\documentclass[lettersize,journal]{IEEEtran}
\usepackage{amsmath,amsfonts}
\usepackage{algorithmic}
\usepackage{algorithm}
\usepackage{array}
\usepackage[caption=false,font=normalsize,labelfont=sf,textfont=sf]{subfig}
\usepackage{textcomp}
\usepackage{stfloats}
\usepackage{url}
\usepackage{verbatim}
\usepackage{graphicx}
\usepackage{cite}
\usepackage[caption=false,font=normalsize,labelfont=sf,textfont=sf]{subfig}
\usepackage{textcomp}
\usepackage{stfloats}
\usepackage{url}
\usepackage{verbatim}
\usepackage{graphicx}
\usepackage{balance}
\usepackage{caption}
\usepackage{subcaption}
\usepackage[table,xcdraw]{xcolor}
\usepackage{booktabs}
\usepackage{multirow}

\usepackage{amsmath} 
\usepackage{amssymb}  
\usepackage{amsthm}
\theoremstyle{definition}

\newtheorem{remark}{Remark}

\newcommand{\R}{\mathbb{R}}
\newcommand{\C}{\mathbb{C}}
\newcommand{\K}{\mathcal{K}}

\newcommand{\norm}[1]{\left\lVert#1\right\rVert}    

\begin{document}

\title{Reduced-order Modeling of Modular, Position-dependent Systems with Translating Interfaces}

\author{Robert A. Egelmeers$^{1}$, Lars A.L. Janssen$^{1\star}$, Rob H.B. Fey$^{1}$, Jasper Gerritsen$^{2}$, Nathan van de Wouw$^{1}$
\thanks{$^{1}$ Dynamics \& Control, Department of Mechanical Engineering,  Eindhoven University of Technology, The Netherlands}
\thanks{$^{2}$ ASMPT Center of Competency, Beuningen, The Netherlands}
\thanks{$^{\star}$ Corresponding author (l.a.l.janssen@tue.nl)}}

\maketitle

\begin{abstract}
Many complex mechatronic systems consist of multiple interconnected dynamical subsystems, which are designed, developed, analyzed, and manufactured by multiple independent teams. 
To support such a design approach,  a modular model framework is needed to reduce computational complexity and, at the same time, enable multiple teams to develop and analyze the subsystems in parallel. 
In such a modular framework, the subsystem models are typically interconnected by means of a static interconnection structure. 
However, many complex dynamical systems exhibit position-dependent behavior ( e.g., induced by translating interfaces) which cannot be not captured by such static interconnection models. 
In this paper, a modular model framework is proposed, which allows to construct an interconnected system model, which captures the position-dependent behavior of systems with translating interfaces, such as linear guide rails, through a position-dependent interconnection structure. Additionally, this framework allows to apply model reduction on subsystem level, enabling a more effective reduction approach, tailored to the specific requirements of each subsystem. Furthermore, we show the effectiveness of this framework on an industrial wire bonder. Here, we show that including a position-dependent model of the interconnection structure 1) enables to accurately model the dynamics of a system over the operating range of the system and, 2) modular model reduction methods can be used to obtain a computationally efficient interconnected system model with guaranteed accuracy specifications.
\end{abstract}

\begin{IEEEkeywords}
Complex dynamical systems, Position-dependent dynamics, Interconnected systems, Modular model framework, Modular model order reduction, Robust performance analysis
\end{IEEEkeywords}

\section{Introduction}
\begin{figure}
    \centering
    \includegraphics[scale=1,page=6]{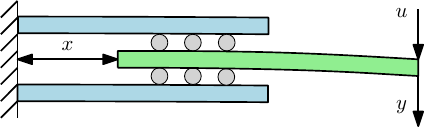} 
    \caption{Example of an interconnected system where the input-to-output dynamic behavior from input $u$ to output $y$ is position-dependent, i.e., it depends on $x$.}
    \label{fig:posdepexample}
\end{figure}

Mechatronic systems often consist of multiple interconnected dynamical subsystems, which are designed individually and in parallel. 
An important sub-class of such complex engineering systems can be modelled as multi-body systems consisting of flexible subsystems translating with respect to each other, in short referred to as systems with translating interfaces.
Examples of these systems are industrial wire bonder machines, printers and high-precision motion stages. 
Accurate, low-order models for the prediction of the motion and structural vibrations of such systems are essential to support model-based (controller) design and to support model-based diagnostics algorithms. 
Obtaining these models is challenged by 1) the large-scale nature of the models, 2) the interconnected nature of these systems, and 3) the position-dependent non-linearity induced by the translating interfaces, which is illustrated with an example in Figure \ref{fig:posdepexample}. The objective of this paper is to provide a modular modeling approach that also supports employing modular complexity reduction techniques, such that, low-order models with guaranteed accuracy specifications can be constructed that are valid on the entire operating range of such systems.

A common engineering approach is to construct a single high-order finite element (FE) model \cite{Hughes2000}, containing all subsystems, to analyze the dynamical behavior of the system as a whole. 
However, to accommodate for increasing performance and accuracy demands, mechatronic systems are becoming increasingly complex.
To enable the design of such complex system models, often, a modular approach is used, where the high-order system model is subdivided into multiple subsystem models that are interconnected through an interconnection structure. 
This enables the development and analysis of the subsystems in parallel by specialized design teams, before integrating them in the interconnected system design. 
In addition, such modular models also allow the reduction of the (computational) complexity of the interconnected system model.
Such a modular modeling approach is commonly used in the systems and control field, e.g., \cite{sandberg2009model,vaz1990}. 
Furthermore, in the structural dynamics field, dynamic substructuring techniques \cite{craig2000coupling,rixen2004dual} and frequency-based substructuring methods \cite{de2008general} are also well-known modular modeling approaches.

Nowadays, such structural dynamics models of complex systems may contain millions of degrees of freedom (DOF), which makes their usage computationally infeasible. 
Therefore, the application of model order reduction (MOR) techniques is necessary to reduce the computational costs of evaluating the dynamical characteristics of these systems,  related to specific input-output combinations and often in a certain frequency range of interest. 
Commonly used MOR techniques include for instance balancing methods \cite{gugercin2004}, moment matching techniques \cite{antoulas2005,Lanczos1950,Arnoldi1951}, and Component Mode Synthesis (CMS) methods \cite{craig1968coupling,herting1985general,rubin1975improved}. 

There exist two main approaches to reduce the complexity (order) of interconnected system models.  
First, MOR may be applied on the level of the interconnected system as a whole.
In contrast, in a modular MOR approach, the MOR of each subsystem model is obtained individually without modifying the interconnection structure after which the subsystem reduced-order models (ROMs) are interconnected. 
In contrast to the interconnected system level approach, a modular approach allows to apply different reduction techniques to each individual subsystem, tailored to the specific requirements of each subsystem \cite{reis2008survey}. Furthermore, the modular approach allows to make changes to individual subsystems without having to reconstruct the complete reduced-order model of the interconnected system. In this case, only the ROMs of the involved subsystems need to be reconstructed. 
Finally, reduction on the level of interconnected system as a whole also typically does not preserve the interconnection structure. 
To address this challenge, structure-preserving methods  have been developed \cite{sandberg2009model,vandendorpe2008,lutowska2012,POORT20234240}.
Furthermore, a modular MOR approach provides a significant computational advantage, as it is more efficient to apply MOR to multiple smaller models compared to applying MOR on a single high-order interconnected model. 

However, for modular MOR, predicting how the errors, introduced by the reduction of the subsystems, propagate through the ROM of the interconnected system as a whole is not trivial. Therefore, if a subsystems models are reduced without considering the effect of subsystem modeling errors on the interconnected ROM, these errors may cause given frequency-response function (FRF) accuracy requirements of the interconnected ROM to be violated \cite{Yeung2011}. 
In \cite{janssen2023modular}, a mathematical framework is presented that enables to relate given FRF accuracy requirements of the ROM of the interconnected system to the FRF accuracy requirements of the subsystem ROMs. 
This approach guarantees satisfaction of the required maximum error bounds on the FRFs of the ROM of the interconnected system based on the accuracy requirements on the FRFs of the reduced subsystem models. To achieve these guarantees, a robust performance analysis approach is used \cite{janssen2022modular}. 
A necessary prerequisite for such approaches is the availability of a high-fidelity, modular and linear model accurately describing the (structural) dynamics of the system.

In many (mechatronic) applications, however, the input-to-output behavior is typically position-dependent, i.e., nonlinear, because their functionality requires that modules of the system translate with respect to each other. 
Examples of such systems include industrial wire bonders, CNC machines and high-precision motion stages. 
Models incorporating the position-dependent behavior of the system are crucial for the purpose of designing the mechatronic system such that it performs according to specifications over the entire operating range of the system \cite{dasilva2008design,dasilva2009integrated}. 
A common approach is to construct linearized models at certain operating points. 
These models can, however, only represent the dynamics locally at a certain position. 
Namely, if the subsystem positions change with respect to each other, then the subsystem models and the interface models need to be connected at different locations. 
This has immediate consequences for the (modular) complexity reduction of the subsystem models, since these typically highly depend on the input- and output-variables with which they interconnect to other subsystem models and the interface models. 
These challenges typically make the modeling (and complexity reduction) tasks highly \textit{time consuming and computationally expensive}.
\begin{figure}
    \centering
    \includegraphics[scale=.8,page=3]{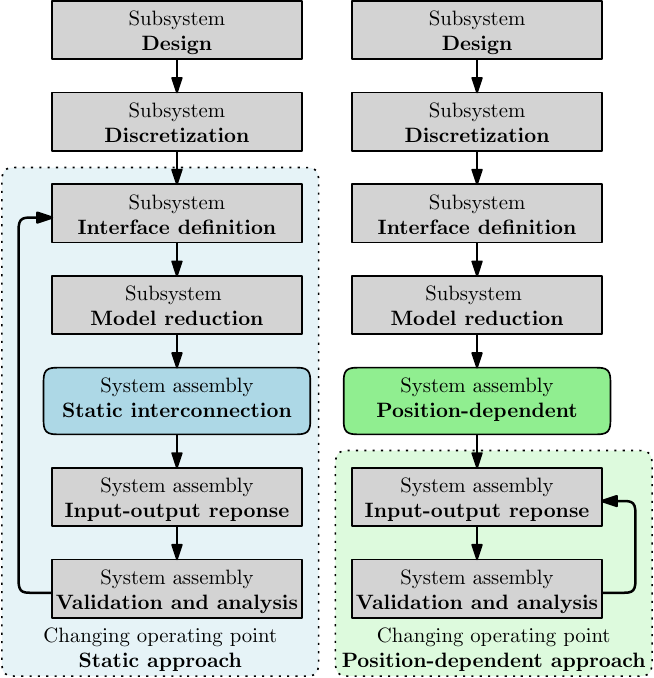} 
    \caption{Comparison of modeling workflows for position-dependent system models with, in the left column, the standard static interconnection structure. In the right column, the proposed position-dependent interconnection structure is given, which significantly reduces required effort for changing the operating points.}
    \label{fig:workflow}
\end{figure}

As a solution, existing methods to model position-dependent behavior of specific complex dynamic systems more efficiently include a position-dependent dynamic substructuring method for machine tools specifically \cite{law2015frequency,LIU2022409}. 
However, there is a need for general methods to construct reduced-order models of position-dependent interconnected systems with \textit{translating} interfaces, such that, firstly, the models can accurately describe the dynamics of systems in the entire operating range of the systems and, secondly, FRF accuracy requirements of the interconnected system ROMs are guaranteed. 

The main contribution of this paper is the development of a general modular model framework for the structural dynamics of systems with translating interfaces (such as, e.g., linear guide rails), that enables to incorporate a position-dependent interconnection structure. 
This is accomplished by introducing fixed grids of virtual interconnection points on the interfaces between subsystems, to be able to connect physical inputs and outputs.
Physical characteristics (such as for example stiffness), specific to each interconnection, are interpolated between two adjacent virtual interconnection points on each interface, depending on their relative position to the physical interconnection. 
The improved modeling workflow in this framework, for systems that exhibit position-dependent behavior, is illustrated in Figure \ref{fig:workflow}. Here, a single modular ROM of the interconnected system can be constructed, that is position-dependent through the interconnection structure. 
This eliminates the need to remodel and reapply MOR on the subsystem models when evaluating different operating points, saving a significant amount of manual and computational effort. 

Moreover, as an additional contribution of this paper, this position-dependent modeling approach is used in combination with the top-down modular MOR framework, described in \cite{janssen2023modular}, which also relies on a modular model framework where the subsystem models are interconnected through an interconnection structure. 
The method developed in \cite{janssen2023modular} guarantees satisfaction of required maximum error bounds on the FRFs of the ROM of the interconnected system based on accuracy requirements on the FRFs of the reduced subsystem models. 
This leaves the position-dependent interconnection structure intact, such that the ROM of the interconnected system can be used to evaluate all operating points of interest. 
As a result, we provide a computationally efficient modular MOR framework for interconnected systems with position-dependent dynamics.

Finally, to evaluate the effectiveness of the proposed framework on real-world applications, a case study on an industrial wire bonder machine (WBM) is presented in this paper. 
A WBM, consisting of multiple modules/stages, makes wired interconnections between a semiconductor die and its packaging. 
In this case study, we show that including a position-dependent interconnection structure 1) enables to accurately model the dynamics of the WBM over the operating range of the system and, 2) modular model reduction methods guaranteeing assembly FRF accuracy specifications can still be used to obtain a computationally efficient model.

This paper is organized as follows. Section \ref{sec:oldframework} gives the modular model framework of systems interconnected through a static interconnection structure. In Section \ref{sec:newframework}, we show how this framework can be extended with a position-dependent interconnection structure, such that we are able to construct a position-dependent interconnected system model. 
In Section~\ref{sec:mor}, we concisely present the modular model reduction method.
The effectiveness of the proposed position-dependent modeling and modular reduction framework is evaluated by means of a case study on an industrial wire bonder in Section \ref{sec:usecase}. Finally, the conclusions are given in Section \ref{sec:conclusions}.

\section{Modular model framework}\label{sec:oldframework}
In this section, we introduce the modular modeling framework that can be used for the purpose of decreasing computational costs, improving interpretability, and enabling an effective, structure-preserving model reduction approach. To obtain such a modular model, complex high-order (mechanical) systems can be subdivided into $k$, less complex, high-order subsystems $j\in\{1,\dots,k\}$, the dynamics of which can be described as second-order ordinary differential equations (ODE) of the form
\begin{equation}\label{eq:ode}
    M_j\ddot q_j(t)+D_j\dot q_j(t)+K_jq_j(t)=F_j(t),
\end{equation}
where $M_j$, $D_j$, and $K_j$ are the mass, damping and stiffness matrices of the respective subsystems with index $j$. The generalized coordinate and force vectors are denoted by $q_j(t)$ and $F_j(t)$, respectively. If we consider state $x_j(t):=\left[q_j^\intercal (t),\dot q_j^\intercal (t)\right]^\intercal$ and input $u_j(t) := F_j(t)$, we can formulate (\ref{eq:ode}) into (descriptor) state-space form,
\begin{equation}\label{eq:ss}
    \begin{split}
        E_j\dot x_j(t)=A_jx_j(t)+B_ju_j(t)\\
        y_j=C_jx_j(t)+D_{ss,j}u_j(t),
    \end{split}
\end{equation}
of order $n_j$, i.e. $A\in\R^{n_j\times n_j}$, with $n_j$ being the number of states of subsystem $j$, with inputs $u_j$ and outputs $y_j$ of dimensions $m_j$ and $p_j$, respectively, and where
\begin{equation}
    E_j = \begin{bmatrix}
        I&0\\0&M_j
    \end{bmatrix}, \text{ and } A_j=\begin{bmatrix}
        0&I\\-K_j&-D_j
    \end{bmatrix}.
\end{equation}
In the input matrix $B_j$, the DOFs corresponding to the external inputs and interconnections with other subsystems are selected. 
In the output matrix $C_j$, the DOFs corresponding to the external outputs and interconnections with other subsystems are selected, and $D_{ss,j}$ is a direct feedthrough matrix. In the Laplace domain, the transfer functions $G_j(s)$, representing the input-output behavior of (\ref{eq:ss}), are defined according to
\begin{equation}
    G_j(s) = C_j\left(sE_j-A_j\right)^{-1}B_j+D_{ss,j},
\end{equation}
with $s\in\C$ as the Laplace variable. 
The subsystem transfer functions are collected in a block-diagonal transfer function
\begin{equation}\label{eq:Gb}
    G_b(s) = \text{diag}\left(G_1(s),\dots,G_k(s) \right),
\end{equation}
with inputs $u_b = \left[u_1^\intercal,\dots,u_k^\intercal \right]^\intercal$ and outputs $y_b = \left[y_1^\intercal,\dots,y_k^\intercal\right]^\intercal$ of dimensions $m_b:=\sum_{j=1}^km_j$ and $p_b:=\sum_{j=1}^kp_j$, respectively.

\begin{figure}
    \centering
    \includegraphics[scale=1,page=4]{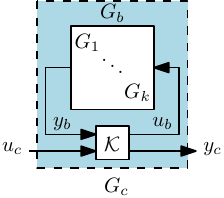} 
    \caption{Block diagram representation of the interconnected system $G_c(s)$ with subsystem models $G_1\dots,G_k$, and the interconnection matrix $\mathcal{K}$.}
    \label{fig:framework}
\end{figure}

Next, the interaction between components is defined in order to derive the model of the interconnected system. One way to achieve this is by using a modular model framework \cite{sandberg2009model,janssen2022}, which is illustrated in Figure \ref{fig:framework}. 
The subsystem models are interconnected according to
\begin{equation}
    \begin{bmatrix}
        u_b\\y_c
    \end{bmatrix} = \mathcal{K}    \begin{bmatrix}
        y_b\\u_c
    \end{bmatrix}:=
    \begin{bmatrix}
        \mathcal{K}_{11}&\mathcal{K}_{12}\\ \mathcal{K}_{21}&\mathcal{K}_{22}
    \end{bmatrix}
    \begin{bmatrix}
        y_b\\u_c
    \end{bmatrix},
\end{equation}
where $u_b$ and $y_b$ are used to connect the subsystems with each other, $u_c$ and $y_c$ are the external $m_c$ inputs and $p_c$ outputs of the interconnected system, and $\mathcal{K}$ is the (static) interconnection matrix.\footnote{In case the external inputs and outputs are directly connected to a subsystem input and output, $u_c$ and $y_c$ contain identical elements in $u_b$ and $y_b$, respectively.} 
The interactions between inputs and outputs of the subsystems are defined in $\mathcal{K}_{11}$. The inputs and outputs of the subsystems, that are chosen as external inputs and outputs, are selected in $\mathcal{K}_{12}$ and $\mathcal{K}_{21}$, respectively. Here, $\mathcal{K}_{22}$ can be seen as a direct feedthrough term from the external inputs to external outputs. Note that in many practical applications, the latter term is zero. The upper linear fractional transformation (LFT) of $G_b(s)$ and $\mathcal{K}$, defined as
\begin{equation}\label{eq:Gc}
    G_c(s) = \mathcal{K}_{21}G_b(s)\left(I-\mathcal{K}_{11}G_b(s) \right)^{-1}\mathcal{K}_{12}+\mathcal{K}_{22},
\end{equation}
defines the transfer function $G_c(s)$ of order $n:=\sum_{j=1}^{k}n_j$ from the external inputs $u_c$ to the external outputs $y_c$ of the interconnected system, see also Figure~\ref{fig:framework}.

\section{Position-dependent modular model framework}\label{sec:newframework}
\begin{figure}
    \centering
    \includegraphics[scale=1,page=2]{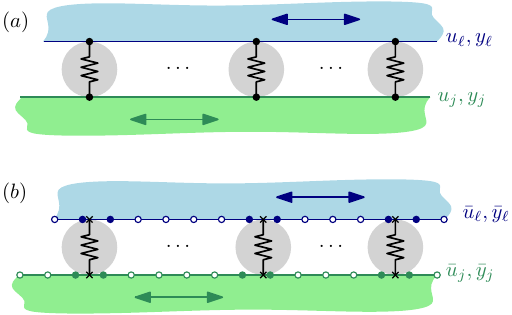} 
    \caption{Illustration of an interconnection structure between two arbitrary subsystems $j$ and $\ell$ with a translating interface with (a) (physical) interconnection points for a specific operation point, and (b) position-dependent approach with a fixed grid of virtual interconnection points. Filled markers indicate virtual interconnections points that are active at a specific operating point, and used to interpolate the characteristics of each interconnection. Empty markers indicate currently inactive interconnection points.}
    \label{fig:int_struct_posdep}
\end{figure}

In many dynamical systems, consisting of multiple interconnected subsystems, the subsystems are interconnected through translating interfaces, such as linear guide rails, to accommodate translation in a single direction. 
The interconnections between the subsystems with these translating interfaces are often represented by translational spring elements \cite{Chang2014AnalyticalAF}, which for instance may model the stiffness of rolling elements in the linear guide rails. 
In the standard framework, as visualized in the left part of Figure~\ref{fig:workflow}, the (inputs and outputs of the) subsystem models $G_j$ need to be reconstructed when the operating point is changed, after which the (momentary) interconnected system is constructed by interconnecting the subsystem models through a static interconnection matrix $\mathcal{K}$, as illustrated in Figure \ref{fig:int_struct_posdep}(a). 
We care to stress once again that this process needs to be repeated every time the operating point is changed (leading to a different $\mathcal{K}$ matrix) which makes such approach computationally prohibitive in practice.

The physical interconnection points are continuously sliding along the interfaces and hence do not always coincide with the input-output pairs of the subsystems models, which are typically defined on discrete (and fixed) grid points at the interfaces. 
Therefore, for this class of position-dependent systems, we propose a novel method to construct the position-dependent model $\bar{G}_c$ (where the bar reflects that position-dependent nature of the model).
To achieve this, we introduce:
\begin{enumerate}
\item Subsystem models $\bar{G}_j$ with fixed, virtual interface points.
\item A position-dependent interconnection matrix $\bar{\mathcal{K}}$.
\end{enumerate}
Here, the key idea is to approximate the characteristics of each interconnection (the points at which the interconnection is active move as the bodies related to the translating interface move with respect to each other) by interpolation of the characteristics of connections, between fixed virtual interconnection points. 

These virtual interconnection points are placed on a pre-defined grid along each substructure interface, as illustrated in Figure \ref{fig:int_struct_posdep}(b).
In this way, the subsystem models $\bar{G}_j$ are fixed for all operating points of the system while the interconnection matrix $\bar{\mathcal{K}}$ is position-dependent.

\subsection{Constructing the position-dependent modular model}
\begin{figure}
    \centering
    \includegraphics[scale=1,page=5]{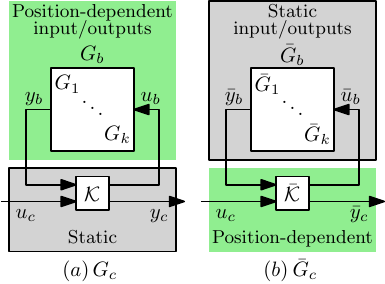} 
    \caption{Block-diagram representation of the interconnected system (a) $G_c$, i.e., with static interconnection matrix $\mathcal{K}$, and (b) $\bar{G}_c$, i.e., with position-dependent interconnection matrix $\bar{\mathcal{K}}$.}
    \label{fig:int_struct_posdep2}
\end{figure}

In the proposed position-dependent framework, the (inputs $\bar{u}_j$ and outputs $\bar{y}_j$ of the) subsystem models $\bar{G}_j$ for all $j\in\{1,\dots,k\}$ remain identical for changing operating points and are connected through a position-dependent interconnection matrix $\bar{\mathcal{K}}$, 
given by
\begin{equation}
    \begin{bmatrix}
        \bar{u}_b\\\bar{y}_c
    \end{bmatrix} = \bar{\mathcal{K}}    \begin{bmatrix}
        \bar{y}_b\\u_c
    \end{bmatrix}:=
    \begin{bmatrix}
        \bar{\mathcal{K}}_{11}&\mathcal{K}_{12}\\ \mathcal{K}_{21}&\mathcal{K}_{22}
    \end{bmatrix}
    \begin{bmatrix}
        \bar{y}_b\\u_c
    \end{bmatrix}.
\end{equation}
Here, the subsystem inputs $\bar{u}_j$ and outputs $\bar{y}_j$ for all $j\in\{1,\dots,k\}$ are combined into $\bar{u}_b$ and $\bar{y}_b$ in the same way as $u_b$ and $y_b$ are constructed.
Note that interconnection matrices $\mathcal{K}_{12}$, $\mathcal{K}_{21}$ and $\mathcal{K}_{22}$ are not affected by the position-dependent framework described in this section. 
Furthermore, a position-dependent transfer function from $u_c$ to $\bar{y}_c$\footnote{The outputs $\bar{y}_c$ and  ${y}_c$ only differ in definition to be able to later compare the difference in the dynamics of the static and position-dependent models.} of the interconnected system $\bar{G}_c$ can be constructed, defined by
\begin{equation}\label{eq:barGc}
    \bar{G}_c(s) = \mathcal{K}_{21}\bar{G}_b(s)\left(I-\bar{\mathcal{K}}_{11}\bar{G}_b(s) \right)^{-1}\mathcal{K}_{12}+\mathcal{K}_{22},
\end{equation}
as illustrated in Figure~\ref{fig:int_struct_posdep2}(b). 

\begin{figure}
    \centering
	\includegraphics[scale=1,page=1]{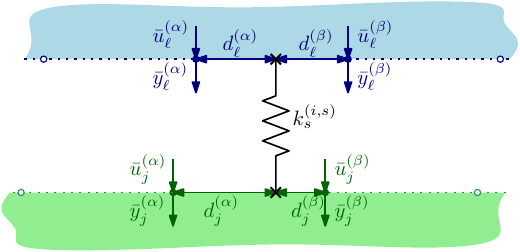} 
    \caption{Representation of a single physical interconnection point, interconnecting two arbitrary subsystems $j$ and $\ell$, with adjacent (i.e., active) virtual interconnection points on each side.}
    \label{fig:single}
\end{figure}

The position-dependent interconnection matrix $\bar{\K}_{11}$ is given by the sum of all interface interconnection matrices, defined by
\begin{align}
\bar{\K}_{11} = \sum_{i=1}^{n_{i}}\bar{\K}_{11}^{(i)}.
\end{align}
Here, the interconnection between two subsystems at each interface $i \in \{1,\dots,n_i\}$, where $n_i$ denotes the total number of interfaces in the system, can be modelled as an interface interconnection matrix $\bar{\K}_{11}^{(i)}$, which is defined by
\begin{align}
\bar{\K}_{11}^{(i)} = \sum_{s=1}^{n_{i,s}}\bar{\K}_{11}^{(i,s)},
\end{align}
i.e., as a summation of the spring interconnection matrix $\bar{\K}_{11}^{(i,s)}$ of each physical spring $s \in \{1,\dots,n_{i,s}\}$ at this interface, where $n_{i,s}$ is the number of springs at interface $i$.
This single spring interconnection matrix $\bar{\K}_{11}^{(i,s)}$ is given by
\begin{align}
\bar{\K}_{11}^{(i,s)} = P_u^{(i,s)}\hat{\K}_{11}^{(i,s)}{P_y^{(i,s)}}^\top \in \R^{m_b \times p_b},
\end{align}
where $P_u^{(i,s)} \in \R^{m_b \times 4}$ and $P_y^{(i,s)} \in \R^{4 \times p_b}$ are permutation matrices that define which virtual interconnection points are active. 
Entry $(f,g)$ of $P_u^{(i,s)}$ is defined by
\begin{align}
P^{(i,s)}_{u,(f,g)} = \begin{cases}
1, & \text{if } g=1 \text{ and } \bar{u}_b^{(f)} \text{ corresponds to } \bar{u}_j^{(\alpha)}\\
1, & \text{if } g=2 \text{ and } \bar{u}_b^{(f)} \text{ corresponds to } \bar{u}_j^{(\beta)}\\
1, & \text{if } g=3 \text{ and } \bar{u}_b^{(f)} \text{ corresponds to } \bar{u}_\ell^{(\alpha)}\\
1, & \text{if } g=4 \text{ and } \bar{u}_b^{(f)} \text{ corresponds to } \bar{u}_\ell^{(\beta)}\\
0, & \text{otherwise.}
\end{cases}
\end{align}
Here, $\bar{u}_j^{(\alpha)}$, $\bar{u}_j^{(\beta)}$, $\bar{u}_\ell^{(\alpha)}$ and $\bar{u}_\ell^{(\beta)}$, are the nearest virtual interface points to the physical spring $s$ on the interconnected subsystems $j\in\{1,\dots,k\}$ and $\ell\in\{1,\dots,k\}$, as illustrated in Figure~\ref{fig:single}, and permutation matrix $P^{(i,s)}_{y}$ is defined equivalently to define the active outputs $\bar{y}_b$. 

Furthermore, the elementary single spring interconnection matrix $\hat{\K}_{11}^{(i,s)}$ is given by
\begin{align}
\label{eq:lin1}
\hat{\K}_{11}^{(i,s)} = Q^{(i,s)} \left[\begin{array}{cc}
-k_s^{(i,s)} & k_s^{(i,s)} \\ k_s^{(i,s)} & -k_s^{(i,s)}
\end{array} \right] Q^{{(i,s)}^\top} \in \R^{4\times 4},
\end{align}
where $k_s^{(i,s)}$ is the translational stiffness of the specific physical spring and
\begin{align}
\label{eq:Q}
Q^{(i,s)} = \left[\begin{array}{cc}
\frac{d_j^{(\alpha)}}{d_j^{(\alpha)}+d_j^{(\beta)}} & 0\\
\frac{d_j^{(\beta)}}{d_j^{(\alpha)}+d_j^{(\beta)}} & 0\\
0 & \frac{d_\ell^{(\alpha)}}{d_\ell^{(\alpha)}+d_\ell^{(\beta)}}\\
0 & \frac{d_\ell^{(\beta)}}{d_\ell^{(\alpha)}+d_\ell^{(\beta)}}
\end{array}\right] \in \R^{4\times 2}.
\end{align}
The linear interpolation matrix $Q^{(i,s)}$ determines how the interconnection between the interface points is distributed, based on the distances to the physical spring locations, given by $d_j^{(\alpha)}$, $d_j^{(\beta)}$, $d_\ell^{(\alpha)}$, and $d_\ell^{(\beta)}$, as illustrated in Figure~\ref{fig:single}. 
With $P$ and $Q$, the position-dependent behavior of the system is described. 
Note that this approach, for a single system position, is similar to typical approaches used when dealing with Finite Element non-conforming meshes \cite{kumar2008finite}.
\begin{remark}
In practice, for different positions of the subsystems, $\bar{\K}_{11}$ is a function of $P^{(i,s)}$ and $Q^{(i,s)}$, as $P^{(i,s)}$ determines which virtual interconnection points are active and $Q^{(i,s)}$ determines the extent in which these points are active. 
These matrices can be obtained automatically, as both $P^{(i,s)}$ and $Q^{(i,s)}$ are determined only by the relative position between the physical and virtual interconnection points.
However, as an arbitrary number of virtual interconnection points can be selected and their positions can also be arbitrarily selected, it is not possible to provide a general position-dependent matrix formulation for $\bar{\K}_{11}$ given any position. 
\end{remark}

Summarizing, with this framework, the position-dependent interconnection matrix $\bar{\K}_{11}$ can approximate the behavior of an arbitrary number of physical springs $n_{i,s}$ on an interface $i$ that are between virtual interface points. With $\bar{\K}_{11}$, the position-dependent interconnected system model $\bar{G}_c$ as in (\ref{eq:barGc}) can be obtained.

\subsection{Accuracy of the position-dependent model}\label{sub4}
As the position-dependent model $\bar{G}_c$ uses linear interpolation (see (\ref{eq:lin1}), (\ref{eq:Q})) between a grid of virtual interconnection points to obtain a model of the physical interconnection between subsystems, a modeling error can be introduced.
The size of this modelling error is dependent on several factors, including:
\begin{itemize}
\item \textbf{The number of virtual interconnection points}: If the number of virtual interconnection points increases, the distance between physical and virtual interconnection points decreases, which will typically lead to a decrease in modeling error. 
However, as already mentioned, increase of the number of virtual interconnection points also leads to an increase in complexity of the model, as the number of (virtual) inputs and outputs of the subsystems increases. This introduces a trade-off between accuracy and complexity of the model. In Section~\ref{sec:usecase}, we will address this trade-off on an industrial use case. 
It will be demonstrated that already with a relatively low number of virtual interconnection points, the error decreases quickly.
\item \textbf{The stiffness of the subsystems}: The ratio between stiffness of the subsystems and the interface stiffness influences the accuracy of the position-dependent modeling approach.
If this ratio is low, the linear interpolation using the virtual points can limit the accuracy of the model. 
If this ratio is high (as in the limit case for rigid bodies), the error introduced by the position-dependent modeling goes to zero. This is demonstrated by a simulation example, given in \ref{appendix}. We show that as the stiffness of the subsystems is increased, the difference in the input-to-output behavior between the models, converges to zero, regardless of the number of virtual interconnection points per interface in the position-dependent model.
\end{itemize}
Another factor influencing the accuracy is the total stroke length between the subsystems, as an increased stroke length generally leads to a higher position-dependency; this would typically also lead to a higher required number of virtual interconnections points.
Finally, high-frequency eigenmodes are more prone to modeling errors using the virtual interconnection points. 
Therefore, it is important to clearly define the frequency range of interest for the model.
In Section~\ref{sec:usecase}, these factors will be taken into account to determine the number and location of the virtual interconnection points on an industrial use case.

\section{Modular model order reduction}
\label{sec:mor}
In this section, we show how to apply model order reduction to reduce the computational costs that are associated with the evaluation of the dynamical characteristics of the (position-dependent) interconnected system, either $G_c(s)$ or $\bar{G}_c(s)$. 

As explained in the introduction, we apply \emph{modular} MOR, i.e., reduction on subsystem level. 
By doing this, the high-order subsystem models, either $G_j(s)$, or $\bar{G}_j(s)$, of order $n_j$ are reduced to the reduced-order subsystem models $\hat{G}_j(s)$ and $\hat{\bar{G}}_j(s)$, respectively, of order $r_j$, where, typically $r_j \ll n_j$. 
This reduction can be done using any MOR method that can achieve accurate subsystem ROMs $\hat{G}_j(s)$.

\begin{figure}
    \centering
    \includegraphics[scale=1,page=7]{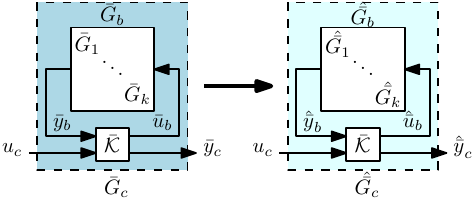} 
    \caption{Block diagram representation of the modular model order reduction approach the reduce the position-depedent interconnected model $\bar{G}_c(s)$ to a reduced-order interconnected model $\hat{\bar{G}}_c(s)$ by reducing the order of the independent subsystem models $\bar{G}_j(s)$, without altering the position-dependent interconnection matrix $\bar{\mathcal{K}}$.}
    \label{fig:red}
\end{figure}

Next, the reduced-order interconnected system models $\hat{G}_c(s)$ or $\hat{\bar{G}}_c(s)$ of order $r:=\sum_{j=1}^{k}r_j$ is constructed by interconnecting the subsystem ROMs $\hat{G}_j(s)$ or $\hat{\bar{G}}_j(s)$, respectively, through the same interconnection matrix as used in the construction of the high-order interconnected system.
For the static $\mathcal{K}$ or the position-dependent $\bar{\mathcal{K}}$ interconnection matrix, the interconnected system ROMs $\hat{G}_c(s)$ and $\hat{\bar{G}}_c(s)$ can then simply be obtained using 
\begin{align}\label{eq:hatGc}
\hat{G}_c(s) &= \mathcal{K}_{21}\hat{G}_b(s)\left(I-{\mathcal{K}}_{11}\hat{G}_b(s) \right)^{-1}\mathcal{K}_{12}+\mathcal{K}_{22}, \text{ or} \\
\hat{\bar{G}}_c(s) &= \mathcal{K}_{21}\hat{\bar{G}}_b(s)\left(I-\bar{\mathcal{K}}_{11}\hat{\bar{G}}_b(s) \right)^{-1}\mathcal{K}_{12}+\mathcal{K}_{22}, 
\end{align} 
respectively. In Figure~\ref{fig:red}, a block-diagram representation of this approach for the proposed position-dependent modelling structure is illustrated.  

To achieve sufficiently accurate models of the reduced interconnected system $\hat{G}_c(s)$, the accuracy of the subsystem ROMs $\hat{G}_j(s)$ needs to be sufficient.
A way to achieve this, is by applying the modular MOR framework, described in \cite{janssen2023modular}, in which FRF accuracy requirements on the ROMs of the subsystem $\hat{G}_j$ are determined, to guarantee FRF error requirements on the ROM of the interconnected system $\hat{G}_c(s)$. 
This method can also be applied to the position-dependent models proposed in this paper to obtain accurate system ROMs $\hat{\bar{G}}_c(s)$.
However, in this case, subsystem ROM accuracy requirements need to be validated in multiple positions of the system, as will be demonstrated in Section~\ref{sec:usecase}. 

The method proposed in~\cite{janssen2023modular} relies heavily on the computation of FRFs of the high-order interconnected system. 
Computing this can be computationally expensive, especially when this needs to be done for many locations in the operating range of the position-dependent system.
However, with the position-dependent modelling approach proposed in this paper, FRFs of the system can be cost-efficiently computed in the entire operating range of the system.
Namely, computation of the FRF matrices is required for all subsystems $\bar{G}_j(s)$ only once because the FRFs of the subsystem models remain identical for for every operating point.
For every operating point, the position-dependent interconnection matrix $\bar{\K}_{11}$ simply needs to be obtained as described in Section~\ref{sec:newframework}.
Then, the FRF matrices of the interconnected system can be determined according to (\ref{eq:barGc})  using computationally cheap matrix computations.

In the next section, we demonstrate that the position-dependent modelling framework can be used to efficiently generate accurate linear reduced-order models of an industrial wire bonder system over the entire operating range.

\section{Use case of an industrial wire bonder machine}\label{sec:usecase}
To illustrate the effectiveness of the proposed framework to approximate the position-dependent dynamics of an interconnected system using a single modular model, the proposed position-dependent framework is applied to a 3D-model of an wire bonder machine, which is illustrated in Figure \ref{fig:3DWBM}. 
For the purpose of achieving a high throughput and positioning-accuracy in the sub-micrometer range at the point of interest, it is crucial to predict the input-to-output behavior of the WBM's x-,  y- and z-motion both fast and accurately. An accurate, low-complexity model for the WBM is essential to support model-based design and to support the design and online operation of control and diagnostic algorithms. Practice has shown, that the dynamic input-to-output behavior of the WBM depends on the relative positions of the flexible subsystems. Therefore, it is essential to construct a (modular) model of the WBM that accurately describes the changing dynamic behavior at different operating points of the stages. 

\subsection{Modeling of the wire bonder machine}
\begin{figure}
    \centering
    \includegraphics[scale=1,page=8]{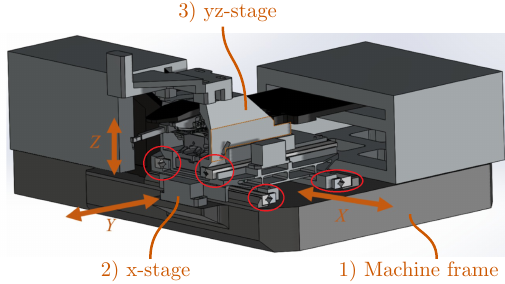} 
    \caption{(Simplified) 3D CAD model of the AB383 wire bonder from ASMPT. The locations of the cross-roller bearing rails are encircled in red.}
    \label{fig:3DWBM}
\end{figure}

In this model, the wire bonder machine consists of three subsystems/modules: 1) the machine-frame, which is rigidly attached to the fixed-world at eight mounting points, 2) the x-stage, that is supported by the machine-frame using two cross-roller bearing rails (modeled by a cross-pattern of translational springs), which enables motion in the x-direction, and 3) the yz-stage, that is supported by the x-stage using two similar cross-roller bearing rails, which enables motion in the y-direction. 
The modules are indicated in Figure~\ref{fig:3DWBM}.
Furthermore, we consider three external (force) inputs and (displacement) outputs for this system. The three force inputs, which define $u_c$, are located at the positions of the motors, actuating, respectively, the x-, y- and z-direction. Actuation in the z-direction is realized by using an elastic rotational hinge around the x-axis. This hinge couples the y- and z-stage (together forming the yz-stage). It should be noted that in our case, we assume that rotations of the z-stage remain small enough to justify the assumption of linear behavior. The three displacement outputs, which define $y_c$, are measured at the encoder locations, enabling to monitor the motions of the x-, y- and z-stage. 

\begin{figure}
    \centering
    \includegraphics[width=0.4\textwidth]{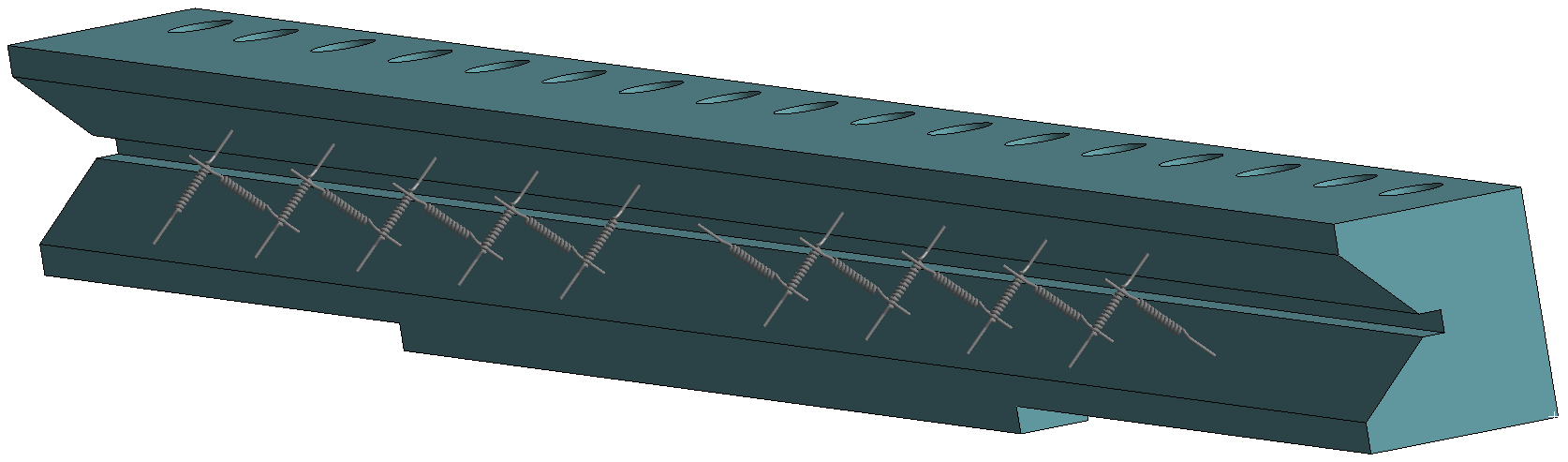}
    \caption{CAD model of one-half of a cross-roller bearing rail, wherein 18 translational springs (indicated by the 18 purple lines) represent the roller bearings.}
    \label{fig:rail}
\end{figure}

\begin{figure}
    \centering
    \includegraphics[width=0.4\textwidth]{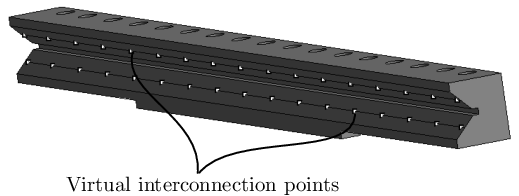}
    \caption{CAD model of one-half of a cross-roller bearing rail, wherein a grid of ($n_v=17$) equidistant virtual interconnection points is defined on each interface.}
    \label{fig:rail_vi}
\end{figure}

In total, there are four cross-roller bearing rails present within the system, each containing nine roller bearings. 
Per rail, there are two pairs of interfaces lying opposite of each other in perpendicular directions. 
To each interface, nine translational springs, representing the roller bearings, are attached, as illustrated in Figure \ref{fig:rail}. Using Ansys \cite{Mechanical}, a grid of equally spaced virtual interconnection points is modeled on each interface, as is illustrated in Figure \ref{fig:rail_vi}, where the total number of virtual interconnection points $n_{v}$ is the same for each interface and varies between $9$, $17$, and $33$ for testing purposes. The DOFs that correspond to the virtual interconnection points, as well as the DOFs that correspond to the external inputs and outputs of the system are defined as interface DOFs.
 
Since the high-order FE-models of the subsystems $\Sigma_j$ (of order $n_j$) consist of millions of DOFs, the subsystem models have to be reduced before exporting their respective (reduced) mass and stiffness matrices, such that they can be handled with a feasible amount of computational cost. 
To achieve this, the CMS reduction method of Craig-Bampton \cite{craig1968coupling} is used, where the DOFs of the system are partitioned into boundary and internal DOFs. Two sets of modes are used.
First, constraint modes (static modes) are defined for all interface DOFs, which were already defined earlier. Second, a set of kept fixed-boundary eigenmodes is defined, which are the kept dynamic modes of the system when all boundary DOFs are fixed. 
Both types of modes form the reduction-basis for this method. 

After reduction, the boundary/interface DOFs are retained, while the internal DOFs are condensed to a (much) smaller set of generalized DOFs. 
The number of kept fixed-boundary eigenmodes, together with the number of boundary/interface DOFs, determines the order $r_j$ of the reduced-order subsystem models. 

For all subsystem models, $200$ fixed-boundary eigenmodes are included in the reduction basis, such that the associated cut-off frequencies $f_{c,j}$, for all subsystems $j=1,2,3$ are multiple times larger than the (normalized \footnote{For confidentiality, all frequencies and FRF magnitudes are normalized.}) largest frequency of interest ($f_i=0.04$ [-]) to ensure that this initial reduction step does not lead to a loss of accuracy in this frequency range. 
 
The specific normalized cut-off frequencies, the number of DOFs before reduction $n_{j,dof}$, and the number of DOFs after reduction $r_{j,dof}$ (for $n_v=9$), for all subsystems are given in Table \ref{table:cutoff}. It is important to note that this initial CMS reduction step in general leads to reduced models which are still too big for some (especially real-time) applications. A second reduction step based on the modular reduction methodology in \cite{janssen2023modular} will be discussed later in Section \ref{sec:usecase_mor}. 
\begin{table}
\footnotesize
\caption{The number of DOFs per subsystem before reduction $n_{j,dof}$ and after reduction $r_{j,dof}$, and the cut-off frequencies $f_{c,j}$ that are associated with the CMS reduction of subsystems $j=1,2,3$.}
\label{table:cutoff}
\centering
\begin{tabular}{|l|l|l|l|}
\hline
$\Sigma_j$ & $1$ & $2$ & $3$ \\ \hline
$f_{c,j}$ [-] & $0.24$ & $0.27$ & $0.13$ \\ \hline
$n_{j,dof}$ & $4.6\cdot10^5$ & $1.6\cdot10^6$ & $3.8\cdot10^6$\\ \hline
$r_{j,dof}$ & $347$ & $431$ & $329$ \\ \hline
\end{tabular}
\end{table}
By using the (reduced) mass and stiffness matrices $M_j$ and $K_j$, respectively, damping matrices $D_j$ for all subsystems $j=1,2,3$ are constructed with $3\%$ modal damping. 
Subsequently, state-space models are constructed for each subsystem using MATLAB \cite{MATLAB}, in which we exploit the sparse nature of the models. Note that all DOFs, associated with the virtual interconnection points, are defined as inputs and outputs in the input and output matrices $B_j$ and $C_j$, respectively, in each subsystem model. 
Using the position-dependent modular model framework, as proposed in Section \ref{sec:newframework}, a position-dependent interconnection matrix $\bar{\mathcal{K}}\in\R^{m_b\times p_b}$ is constructed to interconnect the subsystem models. 
In addition, the external inputs and outputs of the interconnected system are selected. 

Furthermore, for validation purposes, the wire bonder system is also modeled at nine different operating points, using the static interconnection matrix $\mathcal{K}$ from Section \ref{sec:oldframework}, where the x- and y-positions of the x- and yz-motion stages are both varied between $-0.02$, $0$, and $0.02$ m. 
Note that constructing these separate models using the framework from Section \ref{sec:oldframework} is significantly more time consuming than constructing a single position-dependent model, using the proposed methodology. 
In the next section, the position-dependent interconnected system model is compared to these fixed-position models.

\subsection{Comparing $\bar{G}_c$ with $G_c$}
\begin{figure}
    \centering
    \includegraphics[width=0.48\textwidth]{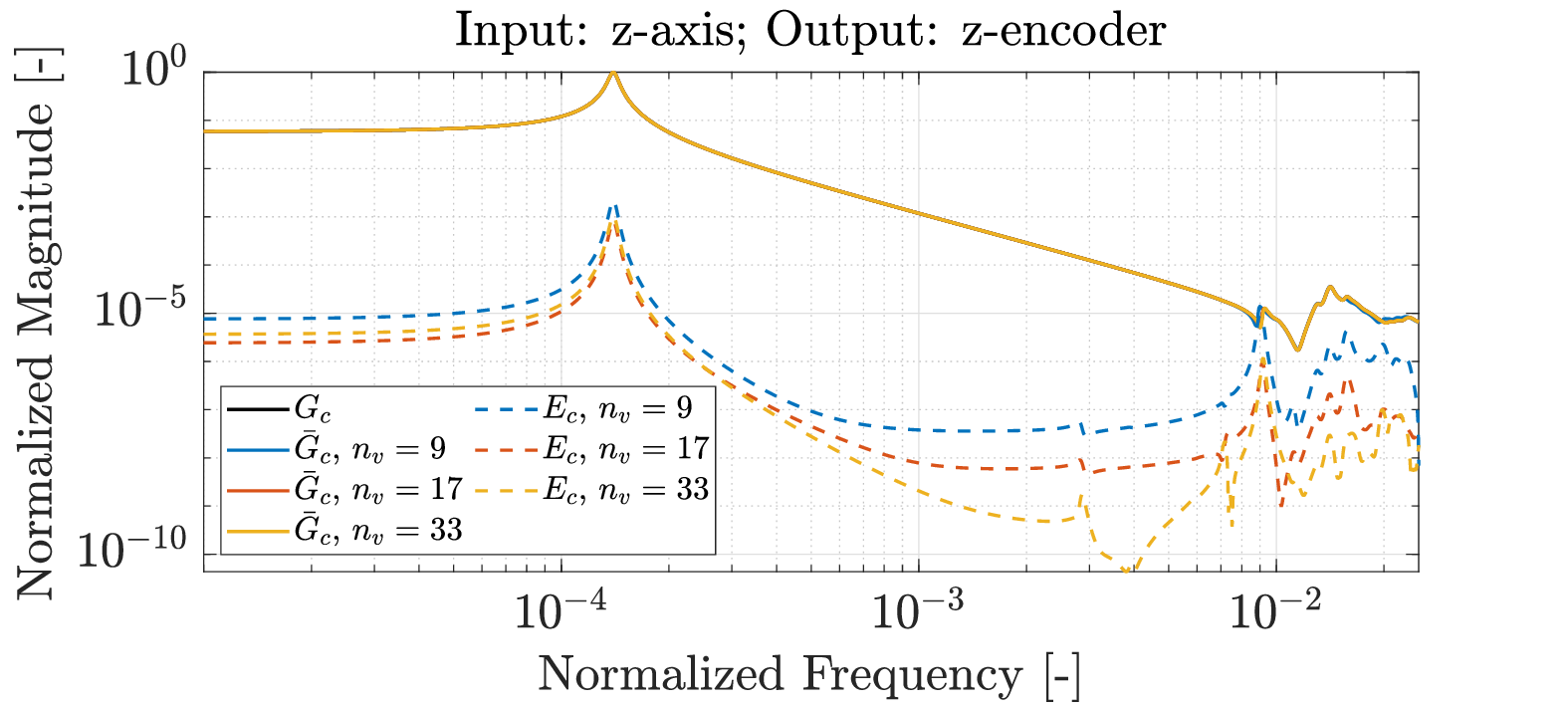}
    \caption{Comparison between the normalized magnitude plots $|G_{c,zz}(i\omega)|$ and $|\bar{G}_{c,zz}(i\omega)|$ for different amounts of virtual interconnection points per interface $n_v$, including the magnitude plots of the corresponding error dynamics $E_{c,zz}(i\omega)$ ($x=0$ [m], $y=0$ [m]).}
    \label{fig:nvcomp33}
\end{figure}
\begin{figure}
    \centering
    \includegraphics[width=0.48\textwidth]{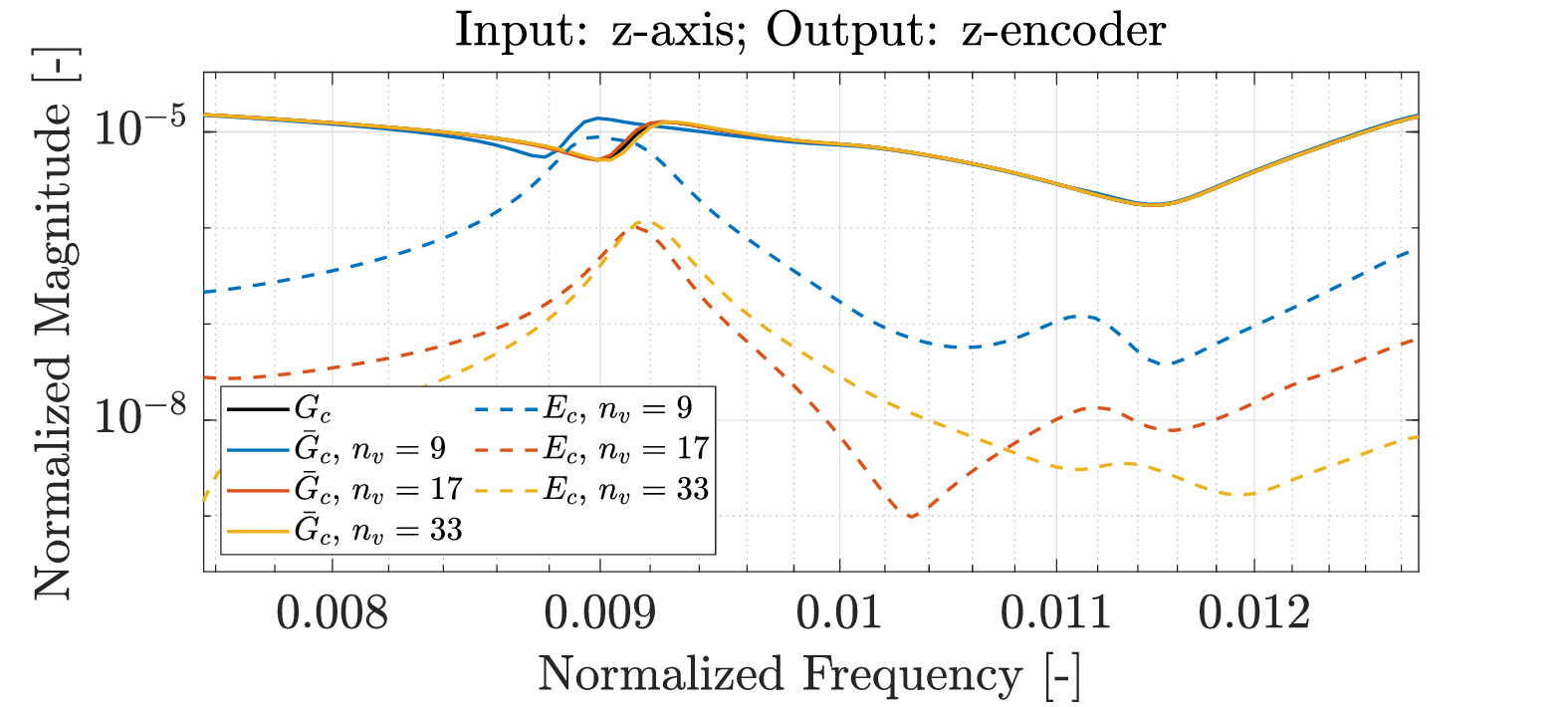}
    \caption{Zoom of Figure \ref{fig:nvcomp33}.}
    \label{fig:nvcomp33z}
\end{figure}
In Figure \ref{fig:nvcomp33}, the normalized magnitude of $G_{c,zz}$ (external z-actuator input to external z-encoder output) of the interconnected system model, constructed with the static interconnection matrix $\mathcal{K}$, is compared to the interconnected system models, constructed with the position-dependent interconnection matrix $\bar{\mathcal{K}}$ for different numbers of virtual interconnection points $n_v$ per interface. 
In addition, a zoom of Figure \ref{fig:nvcomp33} is given in Figure \ref{fig:nvcomp33z} to clearly visualize the differences between the two methods. 
From Figures \ref{fig:nvcomp33} and \ref{fig:nvcomp33z}, it can be observed that the modular modeling approach is able to accurately approximate the dynamics. 

There is a noticeable difference in magnitude between the model $G_{c,zz}$ and the model $\bar{G}_{c,zz}$ for $n_v=9$. 
This difference is significantly less apparent for $n_v=17$ and $n_v=33$. 
By inspecting the error dynamics $E_{c,zz}(s)=G_{c,zz}(s)-\bar{G}_{c,zz}(s)$, it can be observed that if the number of virtual interconnection points per interface $n_v$ is increased, the accuracy of position-dependent model $\bar{G}_{c,zz}(s)$ increases. 
This is expected, because for an increasing number virtual interconnection points per interface, the (interpolation) distances between the physical interconnection point and the adjacent virtual interconnection points become smaller, resulting in a more accurate approximation of the interface dynamics.

\begin{figure}
    \centering
    \includegraphics[width=0.48\textwidth]{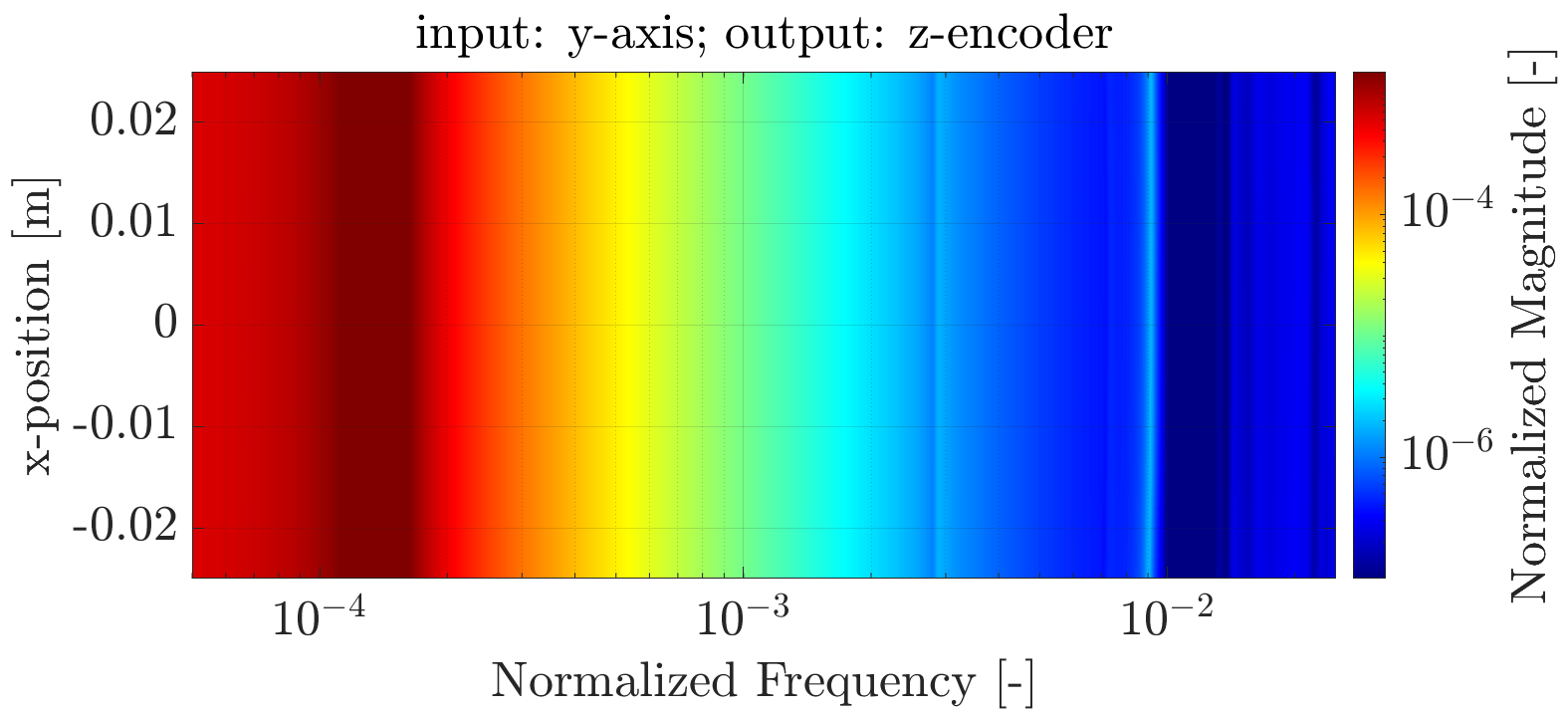}
    \caption{Normalized magnitude plot $|\bar{G}_{c,zy}(i\omega)|$ as a function of a changing operating point in the x-direction ($y=0$ [m]).}
    \label{fig:posdepvarx}
\end{figure}

\begin{figure}
    \centering
    \includegraphics[width=0.48\textwidth]{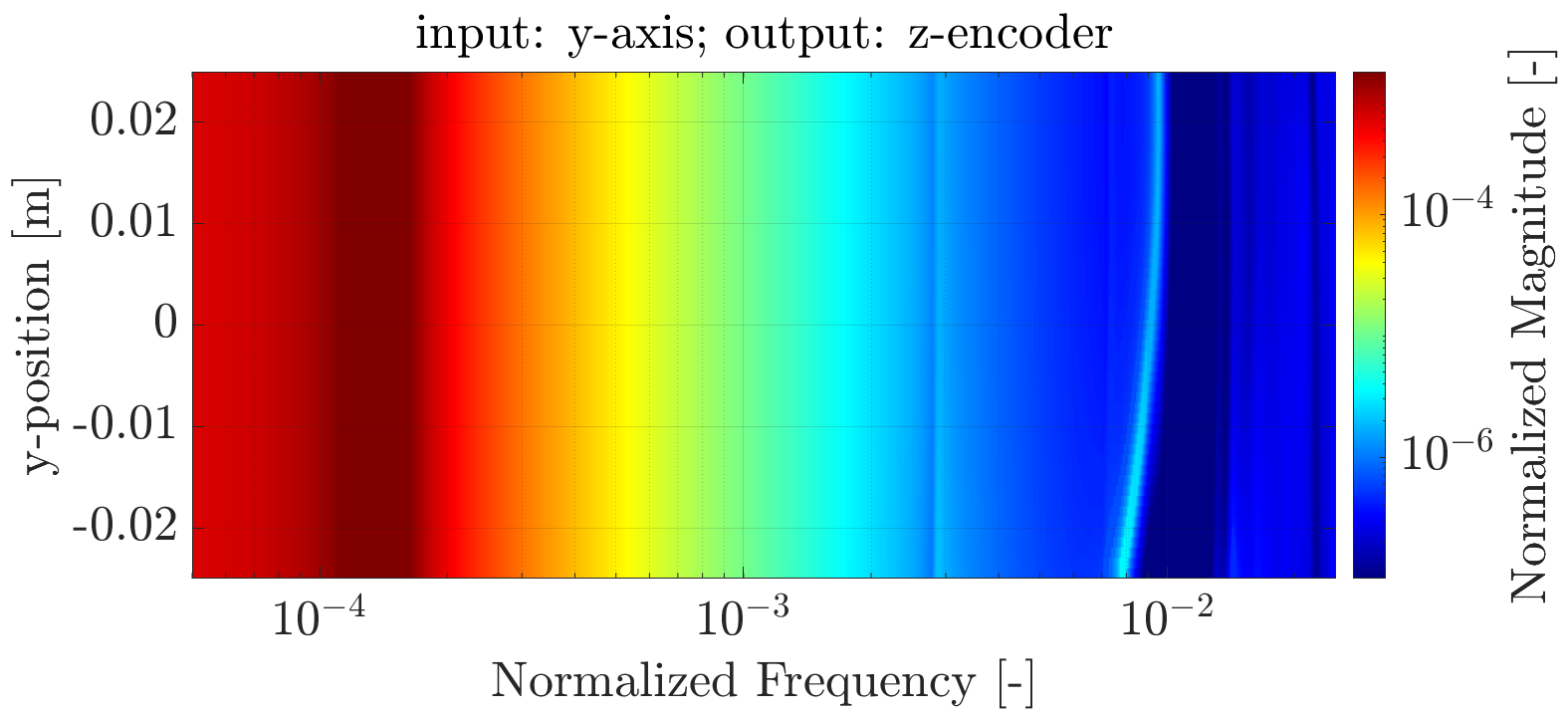}
    \caption{Normalized magnitude plot $|\bar{G}_{c,zy}(i\omega)|$ as a function of a changing operating point in the y-direction ($x=0$ [m]).}
    \label{fig:posdepvary}
\end{figure}

To demonstrate one of the advantages of using the proposed model framework over the static model framework, in Figures \ref{fig:posdepvarx} and \ref{fig:posdepvary}, the normalized magnitudes of the FRFs of $\bar{G}_{c,zy}$ (external y-actuator input to external z-encoder output) are shown as a function of the operating point in the x- and y-direction, respectively. 
Here, a grid of $50$ x- and y-positions are used, respectively. 
It is computational relatively inexpensive to generate these results with the proposed framework, as the FRFs of the subsystems are required to be computed only once. 
In comparison, the conventional approach would require computation of new FRFs from the subsystem models at every operating point.
Then, the FRFs of the interconnected system can be determined for a large number of operating points in fast succession, using cheap matrix operations, as is discussed in Section~\ref{sec:newframework}. 
Figure \ref{fig:posdepvary} also shows that such analysis can quickly visualize how resonance frequencies (e.g., the one near the normalized frequency of $0.01$) depend on the y-position, while Figure \ref{fig:posdepvarx} shows no significant position-dependent behavior.
\begin{figure}
    \centering
    \includegraphics[width=0.48\textwidth]{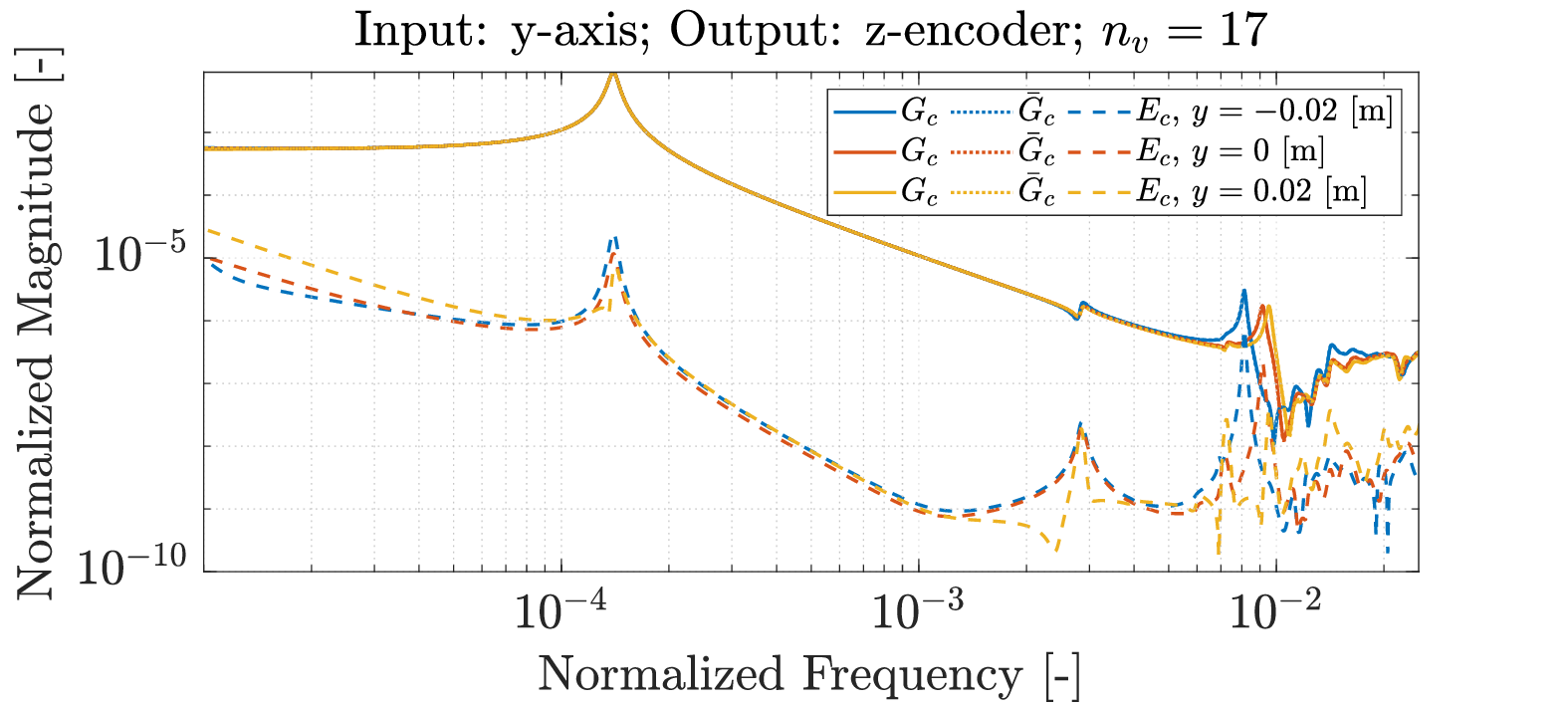}
    \caption{Comparison between the normalized magnitude plots $|G_{c,zy}(i\omega)|$ and $|\bar{G}_{c,zy}(i\omega)|$ for different operating points in the y-direction, including the magnitude plots of the corresponding error dynamics $E_{c,zy}(i\omega)$ ($x=0$ [m]).}
    \label{fig:compy23}
\end{figure}
\begin{figure}
    \centering
    \includegraphics[width=0.48\textwidth]{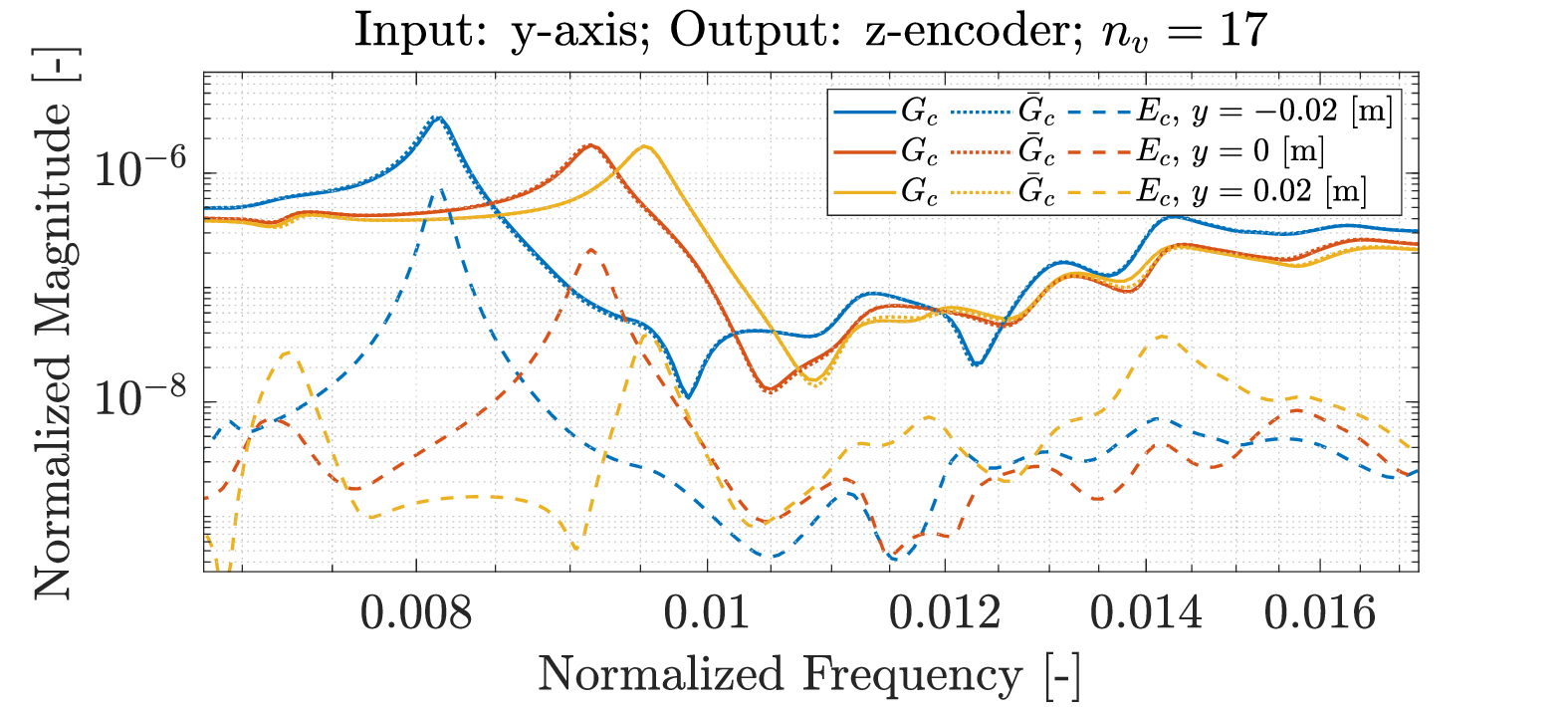}
    \caption{Zoom of Figure \ref{fig:compy23}.}
    \label{fig:compy23z}
\end{figure}
\begin{figure}
    \centering
    \includegraphics[width=0.48\textwidth]{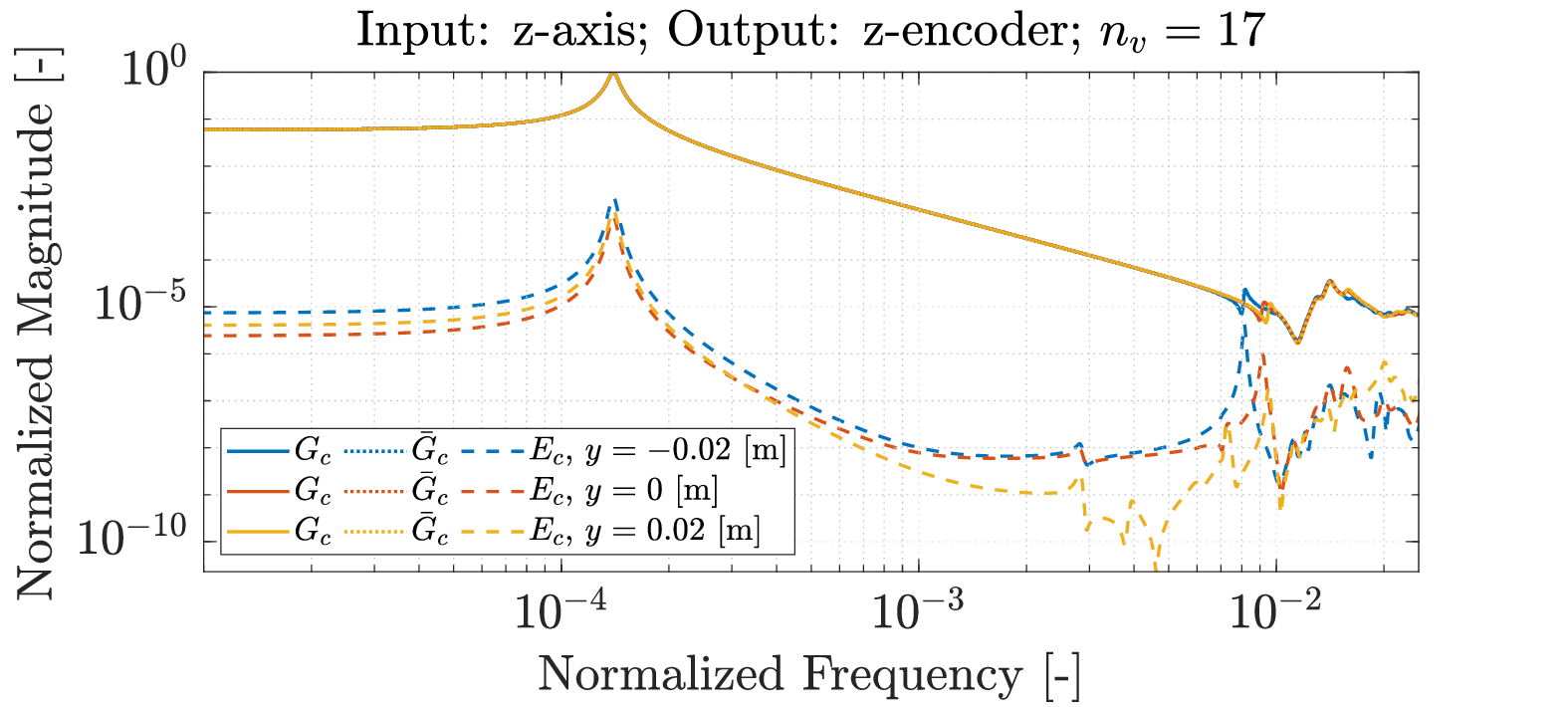}
    \caption{Comparison between the normalized magnitude plots $|G_{c,zz}(i\omega)|$ and $|\bar{G}_{c,zz}(i\omega)|$ for different operating points in the y-direction, including the magnitude plots of the corresponding error dynamics $E_{c,zz}(i\omega)$ ($x=0$ [m]).}
    \label{fig:compy33}
\end{figure}
\begin{figure}
    \centering
    \includegraphics[width=0.48\textwidth]{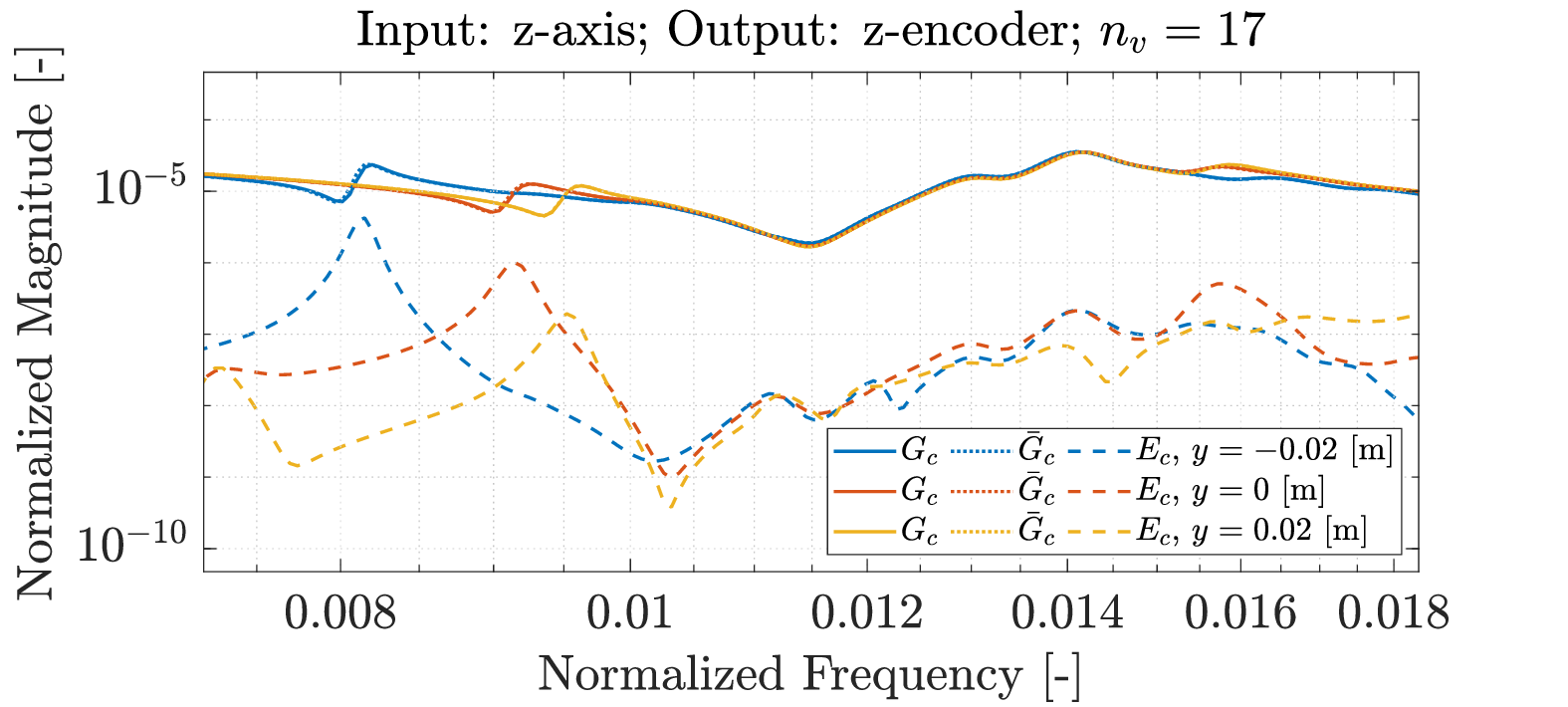}
    \caption{Zoom of Figure \ref{fig:compy33}.}
    \label{fig:compy33z}
\end{figure}

To demonstrate the effectiveness of the position-dependent modular model framework, entries of $G_c(s)$ are compared to similar entries from $\bar{G}_c(s)$ at multiple operating points using $n_v=17$. 
In Figures \ref{fig:compy23} and \ref{fig:compy33}, the normalized magnitudes of the FRFs of both models are plotted along with the normalized magnitudes of the FRFs of the corresponding error dynamics $E_c(s)=G_c(s)-\bar{G}_c(s)$, for the transfer functions from the actuator input $y$ to encoder output $z$, and from actuator input $z$ to encoder output $z$, respectively. 
In addition, Figures \ref{fig:compy23z} and \ref{fig:compy33z} show the same FRFs zoomed in at higher frequencies, where the response of the system is the most sensitive for a changing operating point. 
It can be observed that the wire bonder model shows a clear sensitivity to a changing operating point in the y-direction ($x=0$ [m] in Figures \ref{fig:compy23}-\ref{fig:compy33z}), with regards to its input-to-output behavior, at higher frequencies. 

\begin{figure}
    \centering
    \includegraphics[width=0.48\textwidth]{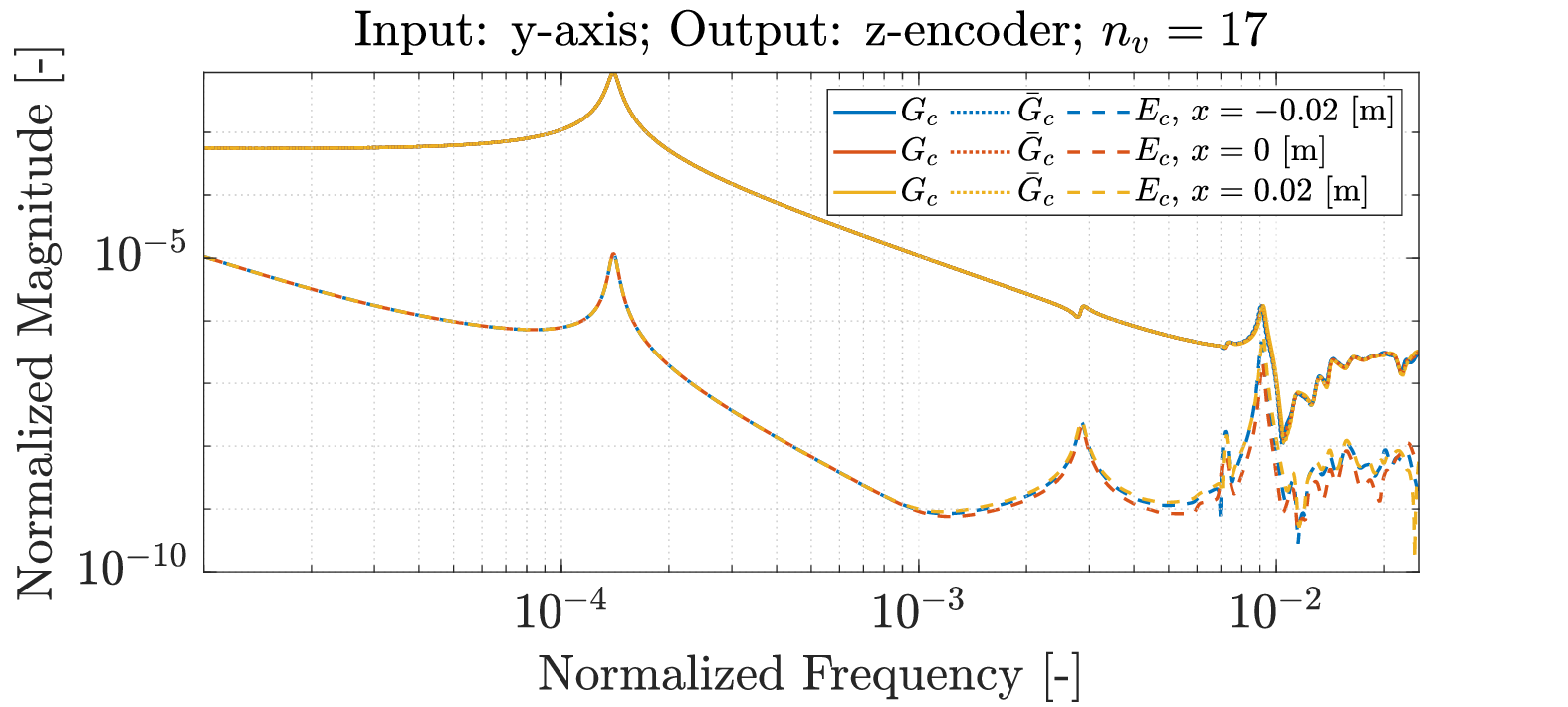}
    \caption{Comparison between the normalized magnitude plots $|G_{c,zy}(i\omega)|$ and $|\bar{G}_{c,zy}(i\omega)|$ for different operating points in the x-direction, including the magnitude plots of the corresponding error dynamics $E_{c,zy}(i\omega)$ ($y=0$ [m]). }
    \label{fig:compx23}
\end{figure}

\begin{figure}
    \centering
    \includegraphics[width=0.48\textwidth]{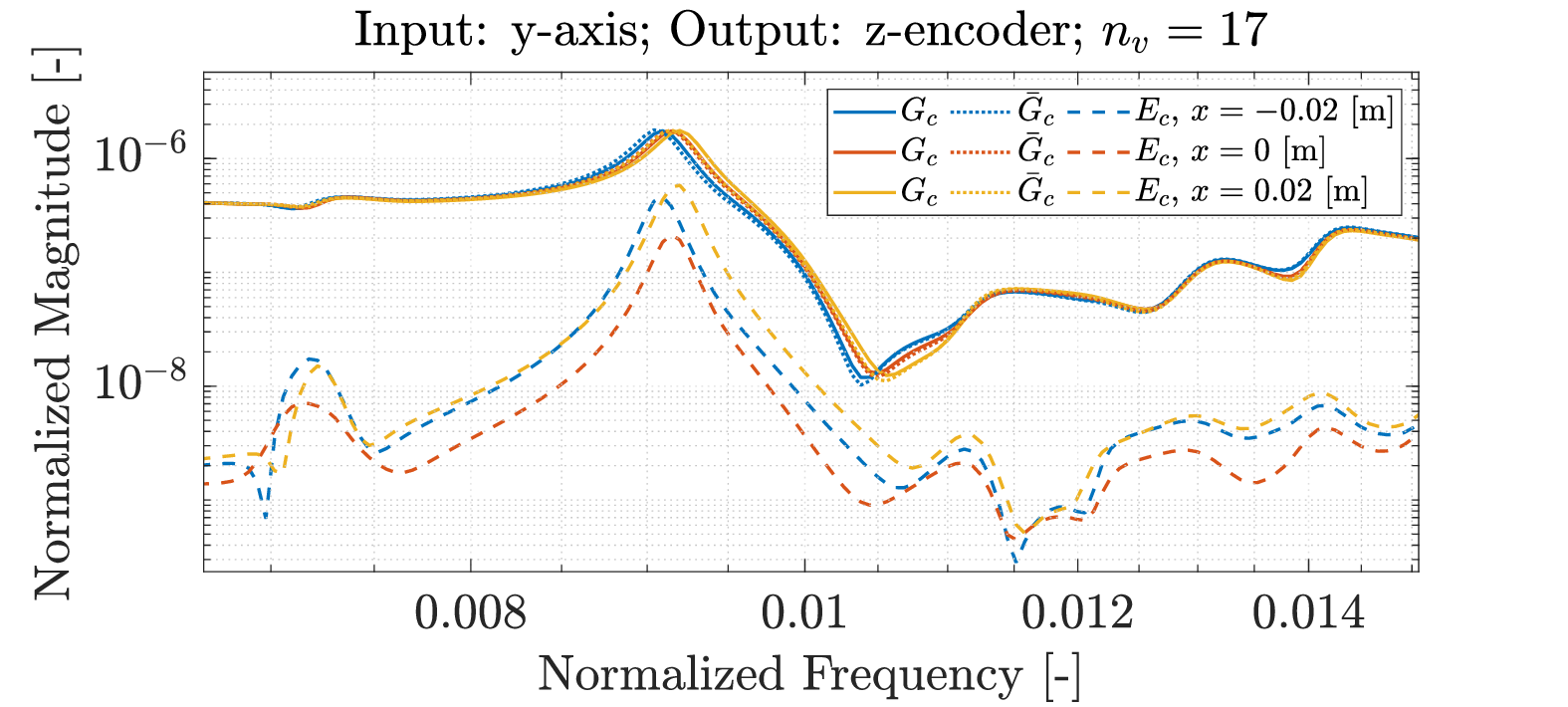}
    \caption{Zoom of Figure \ref{fig:compx23}.}
    \label{fig:compx23z}
\end{figure}

\begin{figure}
    \centering
    \includegraphics[width=0.48\textwidth]{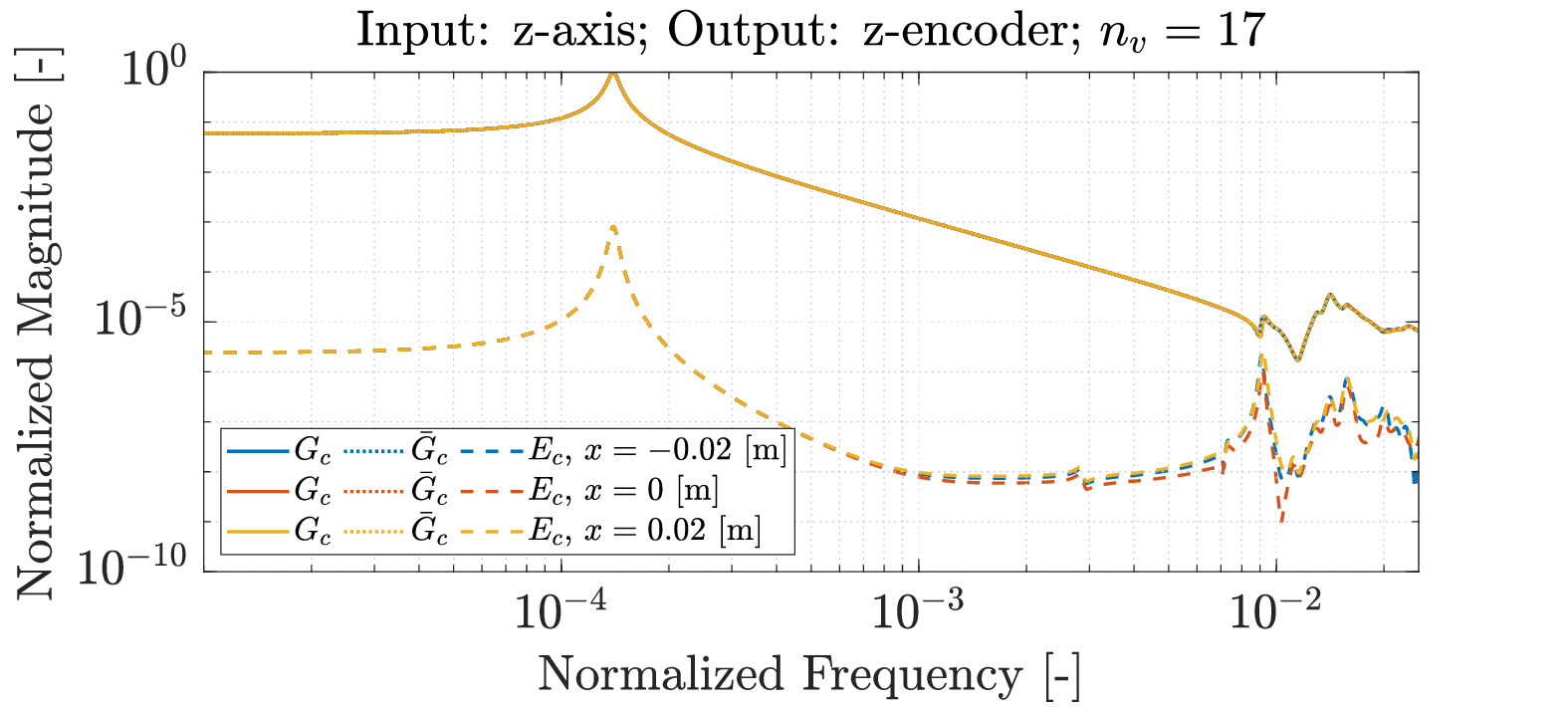}
    \caption{Comparison between the normalized magnitude plots $|G_{c,zz}(i\omega)|$ and $|\bar{G}_{c,zz}(i\omega)|$ for different operating points in the x-direction, including the magnitude plots of the corresponding error dynamics $E_{c,zz}(i\omega)$ ($y=0$ [m]).}
    \label{fig:compx33}
\end{figure}

\begin{figure}
    \centering
    \includegraphics[width=0.48\textwidth]{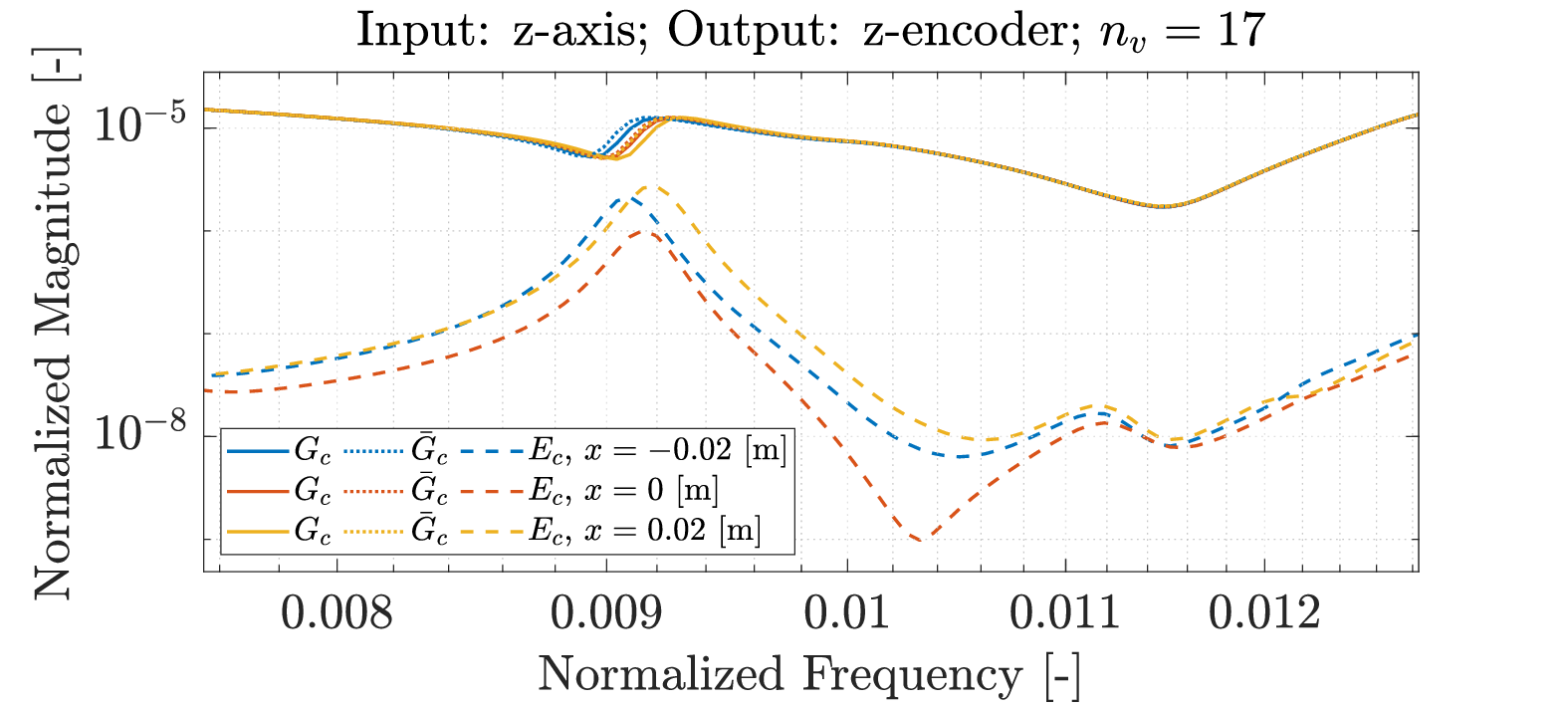}
    \caption{Zoom of Figure \ref{fig:compx33}.}
    \label{fig:compx33z}
\end{figure}

Moreover, it appears that for each operating point, the plotted entries of $G_c(s)$ can be closely approximated by $\bar{G}_c(s)$ with sufficient accuracy. 
In Figures \ref{fig:compx23}-\ref{fig:compx33z}, the normalized magnitudes of the FRFs of $G_c(s)$, $\bar{G}_c(s)$, and $E_c(s)$, are plotted for the same inputs and outputs for multiple operating points in the x-direction ($y=0$ [m]). 
It can be observed that the wire bonder is significantly less sensitive to a changing operating point in the x-direction, with regards to its input-to-output behavior. 
Also, in Figures \ref{fig:compx23}-\ref{fig:compx33z}, the position-dependent interconnected system model $\bar{G}_c(s)$ is able to approximate $G_c(s)$ with sufficient accuracy.


\subsection{Modular model reduction}\label{sec:usecase_mor}
Even though the resulting position-dependent interconnected system model $\bar{G}_c$, after the initial CMS reduction step, is of a sufficiently small order to analyze the input-to-output behavior of the WBM, it is still of a relatively high-order ($n=2166$ for $n_v=9$ virtual interconnection points per interface). 
Therefore, we aim to further reduce the interconnected system model, without compromising too much on model accuracy. 
A modular MOR framework that allows this is recently introduced in \cite{janssen2023modular}. 
This framework enables to determine FRF accuracy requirements on the subsystem ROMs, such that we can guarantee FRF error bounds on the interconnected system ROM. 
Furthermore, this framework allows to use multiple MOR techniques, tailored to the specific requirements on each subsystem. 
This way, only a single ROM needs to be computed for each subsystem, which can be used to obtain accurate system models over the entire operating point of the WBM.
In contrast, the conventional approach would require a dedicated reduction step for each operating point.
This difference is also illistrated in Figure~\ref{fig:workflow}.
In this second reduction step, we use three different MOR techniques: 1) Balanced truncation \cite{antoulas2005}, 2) Craig-Bampton CMS \cite{craig1968coupling}, and 3) Hintz-Herting CMS \cite{herting1985general}. 

Since we made the interconnected system model position-dependent through $\bar{\mathcal{K}}$, it is  a priori unclear how this position-dependency affects the minimal model order that is required, such that a specific FRF accuracy requirement of the ROM $\hat{\bar{G}}_c(s)$ is met at all operating points.
Therefore, we apply the MOR framework from \cite{janssen2023modular} to reduce $\bar{G}_c(s)$ at nine different operating points. 
For each operating point, the minimal subsystem model order $r_j$ is determined for all three reduction techniques, such that the interconnected system model accuracy requirements are based on a maximum on the relative error of $10\%$ up until the largest frequency of interest ($f_i=0.04$ [-]), i.e.,
\begin{equation}
\label{eq:rel_error}
\frac{|\hat{\bar{G}}_{c,zy}(i\omega)-\bar{G}_{c,zy}(i\omega)|}{|\bar{G}_{c,zy}(i\omega)|}<0.1.    
\end{equation}
for all $\omega$ in the frequency range of interest.

Since the MOR framework from \cite{janssen2023modular} does not scale efficiently with large interconnection matrices (it does scale with a high number of system states), we use $n_v=9$ to limit the required computation time, which gives $\bar{\mathcal{K}}\in\R^{308\times308}$. 
This results in a computation time of $116$ minutes per evaluated operating point (AMD(R) Ryzen(TM) 7 5800X3D CPU (4.5Ghz), 64GB RAM). 
The results are presented in Table \ref{table:rom}.\footnote{It should be noted that we can only guarantee the aforementioned accuracy requirement on $\hat{\bar{G}}_c(s)$ for the nine operating points that were evaluated using the reduced subsystem models orders given in Table \ref{table:rom}, which express (for different reduction methods) the minimal subsystem ROM order, such that automatically (\ref{eq:rel_error}) is satisfied).}
\begin{table}
\caption{Minimal subsystem model orders $r_j$ (number of states) to meet the accuracy requirements of the interconnected system model, evaluated at nine different operating points, using three MOR techniques.}
\label{table:rom}
\footnotesize
\centering
\begin{tabular}{l|l|lllllllll}
 &  & \multicolumn{9}{l}{Operating Point} \\ \hline
 &  & \multicolumn{1}{l|}{1} & \multicolumn{1}{l|}{2} & \multicolumn{1}{l|}{3} & \multicolumn{1}{l|}{4} & \multicolumn{1}{l|}{5} & \multicolumn{1}{l|}{6} & \multicolumn{1}{l|}{7} & \multicolumn{1}{l|}{8} & 9 \\ \hline
\multirow{3}{*}{\hspace*{-1.5mm}$\Sigma_1$\hspace*{-1.5mm}} & \hspace*{-1.5mm}BT\hspace*{-1.5mm} & \multicolumn{1}{l|}{118} & \multicolumn{1}{l|}{118} & \multicolumn{1}{l|}{118} & \multicolumn{1}{l|}{118} & \multicolumn{1}{l|}{118} & \multicolumn{1}{l|}{118} & \multicolumn{1}{l|}{118} & \multicolumn{1}{l|}{118} & 118 \\ \cline{2-11} 
 & \hspace*{-1.5mm}CB\hspace*{-1.5mm} & \multicolumn{1}{l|}{162} & \multicolumn{1}{l|}{162} & \multicolumn{1}{l|}{166} & \multicolumn{1}{l|}{166} & \multicolumn{1}{l|}{166} & \multicolumn{1}{l|}{166} & \multicolumn{1}{l|}{178} & \multicolumn{1}{l|}{178} & 170 \\ \cline{2-11} 
 & \hspace*{-1.5mm}HH\hspace*{-1.5mm} & \multicolumn{1}{l|}{162} & \multicolumn{1}{l|}{162} & \multicolumn{1}{l|}{162} & \multicolumn{1}{l|}{162} & \multicolumn{1}{l|}{162} & \multicolumn{1}{l|}{174} & \multicolumn{1}{l|}{170} & \multicolumn{1}{l|}{166} & 166 \\ \hline
\multirow{3}{*}{\hspace*{-1.5mm}$\Sigma_2$\hspace*{-1.5mm}} & \hspace*{-1.5mm}BT\hspace*{-1.5mm} & \multicolumn{1}{l|}{252} & \multicolumn{1}{l|}{252} & \multicolumn{1}{l|}{252} & \multicolumn{1}{l|}{252} & \multicolumn{1}{l|}{252} & \multicolumn{1}{l|}{252} & \multicolumn{1}{l|}{253} & \multicolumn{1}{l|}{253} & 253 \\ \cline{2-11} 
 & \hspace*{-1.5mm}CB\hspace*{-1.5mm} & \multicolumn{1}{l|}{434} & \multicolumn{1}{l|}{434} & \multicolumn{1}{l|}{434} & \multicolumn{1}{l|}{434} & \multicolumn{1}{l|}{434} & \multicolumn{1}{l|}{434} & \multicolumn{1}{l|}{434} & \multicolumn{1}{l|}{434} & 434 \\ \cline{2-11} 
 & \hspace*{-1.5mm}HH\hspace*{-1.5mm} & \multicolumn{1}{l|}{334} & \multicolumn{1}{l|}{334} & \multicolumn{1}{l|}{334} & \multicolumn{1}{l|}{338} & \multicolumn{1}{l|}{334} & \multicolumn{1}{l|}{350} & \multicolumn{1}{l|}{342} & \multicolumn{1}{l|}{342} & 342 \\ \hline
\multirow{3}{*}{\hspace*{-1.5mm}$\Sigma_3$\hspace*{-1.5mm}} & \hspace*{-1.5mm}BT\hspace*{-1.5mm} & \multicolumn{1}{l|}{328} & \multicolumn{1}{l|}{328} & \multicolumn{1}{l|}{328} & \multicolumn{1}{l|}{328} & \multicolumn{1}{l|}{328} & \multicolumn{1}{l|}{328} & \multicolumn{1}{l|}{328} & \multicolumn{1}{l|}{328} & 328 \\ \cline{2-11} 
 & \hspace*{-1.5mm}CB\hspace*{-1.5mm} & \multicolumn{1}{l|}{332} & \multicolumn{1}{l|}{332} & \multicolumn{1}{l|}{332} & \multicolumn{1}{l|}{332} & \multicolumn{1}{l|}{332} & \multicolumn{1}{l|}{332} & \multicolumn{1}{l|}{332} & \multicolumn{1}{l|}{332} & 332 \\ \cline{2-11} 
 & \hspace*{-1.5mm}HH\hspace*{-1.5mm} & \multicolumn{1}{l|}{240} & \multicolumn{1}{l|}{240} & \multicolumn{1}{l|}{240} & \multicolumn{1}{l|}{240} & \multicolumn{1}{l|}{240} & \multicolumn{1}{l|}{240} & \multicolumn{1}{l|}{252} & \multicolumn{1}{l|}{248} & 248
\end{tabular}
\end{table}

From the results, it can be concluded that there is little variation in the minimally required subsystem model orders between the operating points, when using the same reduction technique. 
This indicates that, in this case study, the accuracy requirements of the subsystems are not very sensitive to the operating point of the system. 
Furthermore, it can be concluded that the balanced truncation method is the most effective MOR technique for the first and second subsystem models, whereas the CMS method of Hintz-Herting is the most effective MOR technique for the third subsystem. 
In Table \ref{table:finalrom}, the resulting subsystem model orders, before and after reduction, are presented. 

These results show that the MOR framework, introduced in \cite{janssen2023modular}, in combination with the position-dependent model introduced in this paper allows to construct an accurate, modular, and position-dependent ROM of the interconnected system model that facilitates to obtain FRFs at the operating range of the interconnected system in a fast and accurate manner.
\begin{table}
\caption{Optimal MOR technique and model order $r_j$ per subsystem to meet the accuracy requirements of the interconnected system model at all nine operating points.}
\label{table:finalrom}
\footnotesize
\centering
\begin{tabular}{l|l|l|l|l}
& Method & $n_j$ & $r_j$ & Reduction \% \\ \hline
$\Sigma_1$&BT & 646 & 118 & 80.8 \\ \hline
$\Sigma_2$&BT & 862 & 253 & 70.6 \\ \hline
$\Sigma_3$&HH & 658 & 252 & 61.7 \\ \hline
\textbf{Total:}& & 2166 & 623 & 71.2 \\ 
\end{tabular}
\end{table}
\normalsize
\section{Conclusions}\label{sec:conclusions}
In conclusion, this paper introduces a novel modular model framework for position-dependent dynamical systems consisting of multiple flexible bodies and translating interfaces, while addressing a key limitation in current modeling approaches. 
Namely, the conventional modular model framework requires remodeling of subsystems when evaluating the input-to-output behavior at different operating points (due to the position-dependency), which often has to be done manually and/or is computationally costly. 
The proposed framework overcomes this roadblock by incorporating a position-dependent interconnection structure through fixed grids of virtual interconnection points, introduced along the interfaces. This ensures that the subsystem models are identical at all operating points of the system. 

In addition, by making the interconnection structure position-dependent, the proposed framework enables the construction of a modular reduced-order model for the interconnected system that incorporates position-dependent behavior without sacrificing modularity. 
To obtain this reduced-order model, we use a recently introduced modular model-order reduction framework that ensures that FRF error bounds on the reduced-order model of the interconnected system are satisfied (at a selection of operating points), enhancing the reliability and applicability of the proposed modular model framework.

Finally, the proposed modular modeling framework is applied to an industrial wire bonder model, in which we demonstrate the modeling method. Here, we validate by comparison with the classical modeling approach that the model is accurate when enough virtual interconnection points are used. 
We show, for this case study, an accurate, modular, and position-dependent reduced-order model that can be used to obtain frequency-response functions at the operating range of the system in a fast and accurate manner. 
Such models can subsequently be used to support model-based design guaranteeing performance over the entire operating range.

\section{Acknowledgments}
This publication is part of the project Digital Twin with project number P18-03 of the research programme Perspectief which is (mainly) financed by the Dutch Research Council (NWO).

%
%
%
%

\appendix
\section{Preservation of stiffness properties}\label{appendix}
In this section, we demonstrate that, for a certain operating point, the translational and rotational stiffness (with respect to an arbitrary point on the interface) of the position-dependent interconnection matrix $\bar{\mathcal{K}}_{11}$, is equivalent to the static interconnection matrix $\mathcal{K}_{11}$. Consider the interconnected system $G_c(s)$, interconnected through a static interconnection matrix $\mathcal{K}$, and $\bar{G}_c(s)$, interconnected through a position-dependent interconnection matrix $\bar{\mathcal{K}}$. In the simulation example below we show that when the stiffness of the subsystems is increased, the difference in input-to-output behavior between $G_c(s)$ and $\bar{G}_c(s)$ converges to zero. However, in practice the subsystems are flexible. Therefore, errors are introduced by interpolating the characteristics of the physical interconnections to the inputs and outputs, corresponding to the virtual interconnection points, as described in Section \ref{sec:newframework}.

\begin{figure}
    \centering
    \includegraphics[width=0.48\textwidth]{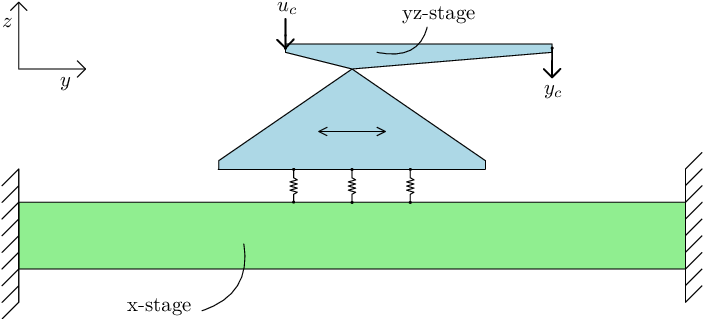}
    \caption{2D wire bonder model, consisting of an x-stage and a yz-stage (with external input $u_c$ and output $y_c$), which are interconnected by three bearings (modeled as translational springs).}
    \label{fig:2dwbm}
\end{figure}

\begin{figure}
    \centering
    \includegraphics[width=0.48\textwidth]{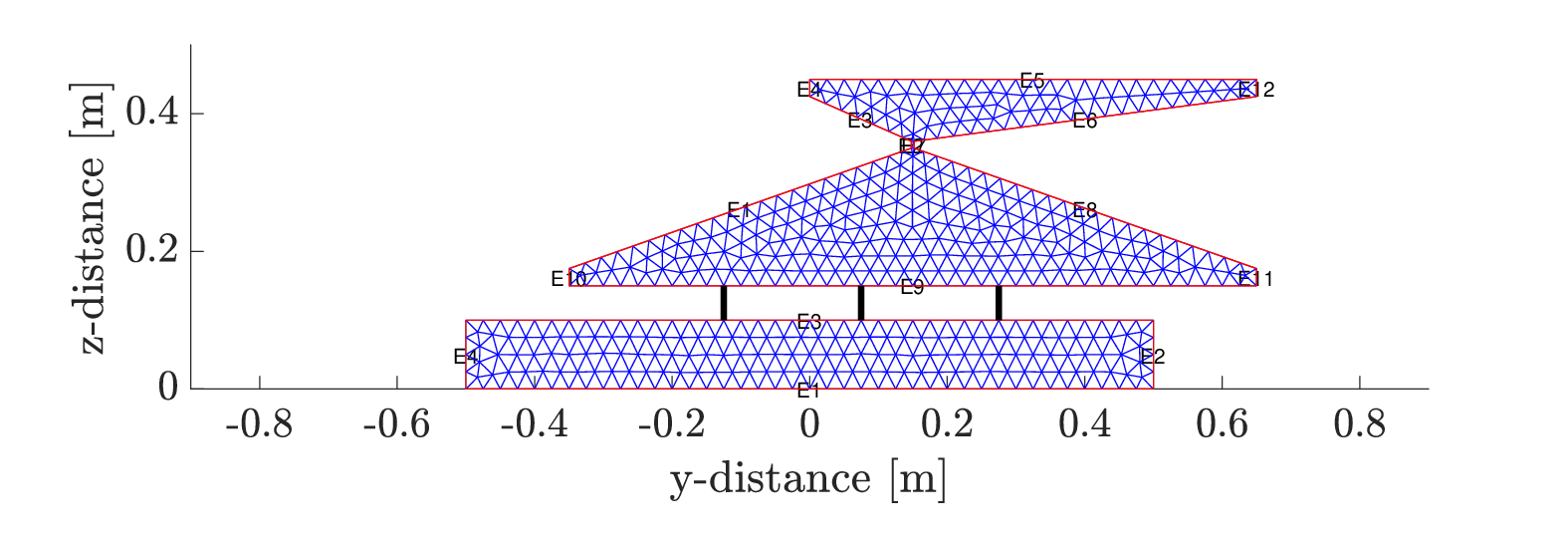}
    \caption{2D FE models of the x-stage and yz-stage of the WBM, interconnected through three translational springs (indicated by black vertical lines between the subsystems).}
    \label{fig:mesh}
\end{figure}

Using a simplified 2D FE-model of the WBM, we demonstrate that, for an increasing subsystem stiffness, the error between the FRFs of $G_c(s)$ and $\bar{G}_c(s)$ converges to zero. The WBM in this model (illustrated in Figure \ref{fig:2dwbm}) consists of two subsystems: 1) the x-stage, which is modeled as a clamped beam, because there are no DOFs in the x-direction, and 2) the yz-stage, that is supported by the x-stage using three bearings (modeled by three vertical translational springs with stiffness $k_s=2\cdot10^6$ [N/m]), which enables motion in the y-direction. Furthermore, the horizontal dimensions of the yz-stage are stretched in this model to exaggerate position-dependent behavior (see Figure \ref{fig:mesh}).

Using MATLAB's \cite{MATLAB} PDE toolbox, FE-models of the subsystems have been generated (see Figure \ref{fig:mesh}), where both are discretized using quadratic triangular plane-stress elements, with a mass density of $7800$ kg/m$^2$, a Poisson's ratio of $0.3$, and $3\%$ modal damping. The Young's modulus is treated as a variable in this example to vary the stiffness of the subsystems. The interconnected system models $G_c(s)$ and $\bar{G}_c(s)$ (with $n_v=5$ virtual interconnection points per interface) have been constructed according to the modular model frameworks from Sections \ref{sec:oldframework} and \ref{sec:newframework}, respectively. 

\begin{figure}
    \centering
    \includegraphics[width=0.48\textwidth]{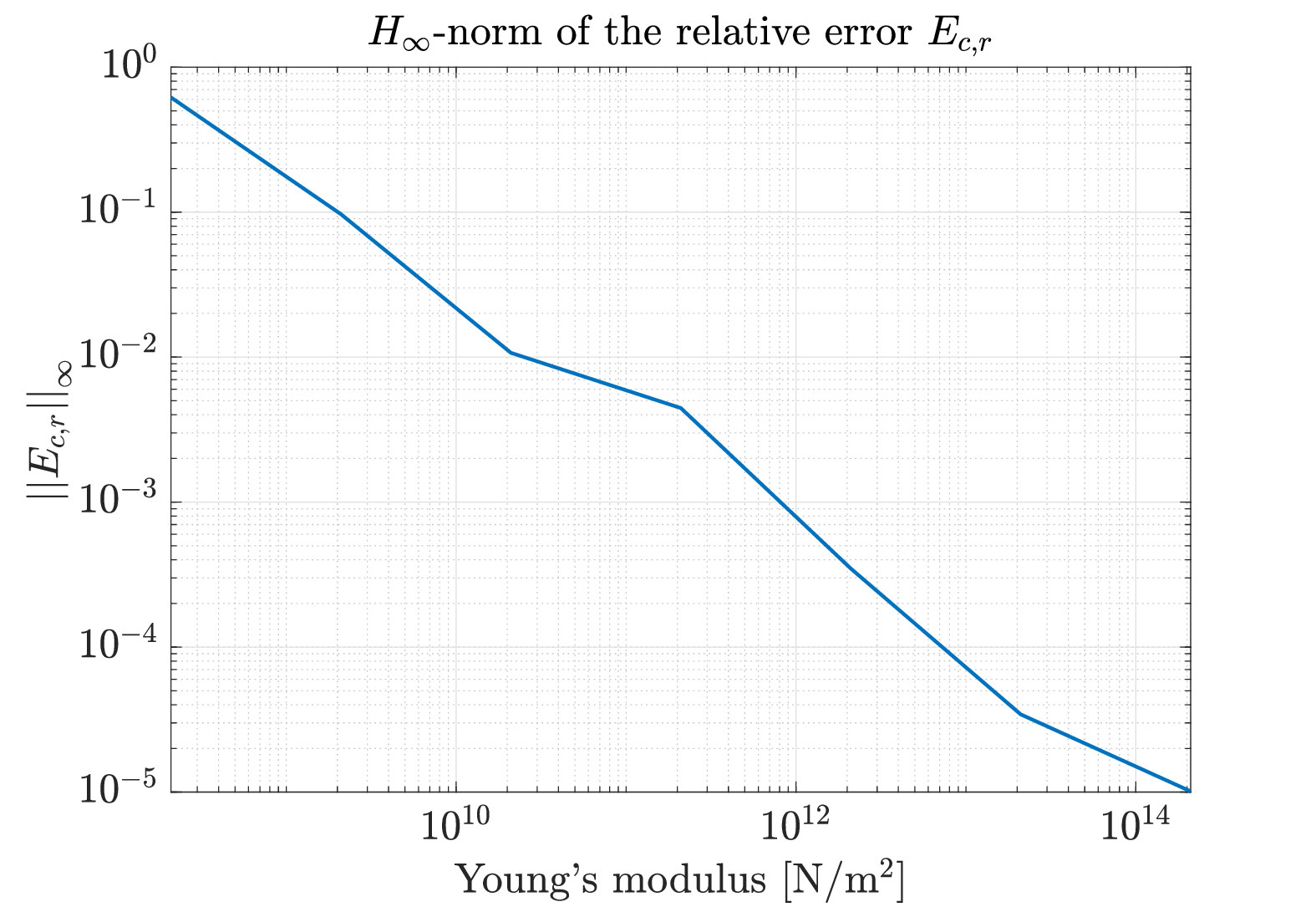}
    \caption{Plot of the $H_\infty$ norm of the relative error FRF $E_{c,r}(i\omega)$, as a function of the Young's modulus of the subsystems.}
    \label{fig:erel}
\end{figure}
To quantify the accuracy of the FRF of $\bar{G}_c(s)$ compared to the FRF of $G_c(s)$, as a function of the subsystem stiffness, the $H_\infty$ norm of the relative error $E_{c,r}(s)$ is calculated according to
\begin{equation}
     \norm{E_{c,r}(s)}_\infty = \norm{\frac{\bar{G}_c(s)-G_c(s)}{G_c(s)}}_\infty = \underset{\omega\in\R}{max}\left|E_{c,r}(i\omega)\right|.  
\end{equation}
The results are plotted in Figure \ref{fig:erel}. It can be observed that the norm of the relative error FRF $E_{c,r}(i\omega)$ converges to zero when the subsystem stiffness is increased. This indicates that when the subsystems behave as rigid bodies, the static and position-dependent interconnection matrices $\mathcal{K}$ and $\bar{\mathcal{K}}$, respectively, have an identical effect on the input-to-output behavior of the interconnected system. This implies that the stiffness, described by the position-dependent interconnection structure, is preserved in the position-dependent model framework, described in Section \ref{sec:newframework}. In real-world applications, the subsystems do not behave as rigid-bodies, and, therefore, the error between $G_c(s)$ and $\bar{G}_c(s)$ can only be decreased by adding more virtual interconnection points on each interface (or by placing them more efficiently).
\end{document}